\documentclass[aps,prb,twocolumn,showpacs,superscriptaddress]{revtex4}

\usepackage{wasysym}
\usepackage{ifsym}

\usepackage{amsfonts}
\usepackage{mathrsfs}

\usepackage{amsthm}
\usepackage{epsfig}
\usepackage{graphicx}
\usepackage{url}
\usepackage{hyperref}
\usepackage{float}
\usepackage{pstricks}
\usepackage{color}
\usepackage{tipa}
\usepackage{subfig}
\usepackage{xcolor}
\usepackage{mathrsfs}
\theoremstyle{plain}

\newtheorem{defn}{Definition}[section]
\newtheorem{thm}{Theorem}[section]
\newtheorem*{pf}{Proof}

\newtheorem{ex}{Example}[section]

\begin{document}

\title{Axion field theory approach and the classification of interacting topological superconductors}
\author{Yingfei Gu}
\author{Xiao-Liang Qi}
\affiliation{Department of Physics, Stanford University, Stanford, CA 94305, USA}
\date{\today}

\begin{abstract}

In this paper, we discuss the topological classification of time-reversal invariant topological superconductors. Based on the axion field theory developed in a previous work (Phys. Rev. B {\bf 87} 134519 (2013)), we show how a simple quantum anomaly in vortex-crossing process predicts a $\mathbb{Z}_{16}$ classification of interacting topological superconductors, in consistency with other approaches. We also provide a general definition of the quantum anomaly and a general geometric argument that explains the $\mathbb{Z}_{16}$ on more general grounds. Furthermore, we generalize our approach to all $4n$ dimensions (with $n$ an integer), and compare our results with other approaches to the topological classification.

\end{abstract}
\maketitle

\section{Introduction}

Topological states of matter have attracted enormous attention since the discovery of integer and fractional quantum Hall effects in 1980's\cite{klitzing1980new,tsui1982two}. The robust and accurate quantization of Hall conductance was soon realized to be a topological property of this quantum matter\cite{laughlin1981,thouless1982quantized}. The topological properties of quantum Hall systems have been related to topological field theories known as the Chern-Simons theory\cite{zhang1989effective,jackiw1984fractional,witten1989quantum,wen1991non} 

More recently, new topological quantum matters have been predicted and experimentally realized in the past decade, known as topological insulators (TI) and topological superconductors (TSC)\cite{moore2010birth,qi2011topological,hasan2010colloquium,qi2010quantum}.
 The understanding of TI and TSC has been generalized to a more generic theory of symmetry protected topological states(SPTs)\cite{chen2012symmetry,senthil2014symmetry}, which clarifies the interplay of symmetries and topological properties in quantum states of matter. Among SPTs, the specific systems we will focus on in this paper are the time-reversal invariant (TRI) topological superconductors (known as the DIII class) in $d+1=4n$ dimensions.\cite{roy2008topological,qi2009time,schnyder2008classification}

Topological superconductors are gapped superconductors with topologically robust gapless quasi-particles propagating on the boundary\cite{schnyder2008classification,roy2008topological,qi2009time}. 
The original proposals of topological superconductors are based on the topological classification of quasiparticle band structures in a BCS mean-field theory. In $(3+1)$-d, the topological classification is given by an integer $\nu$, which corresponds to the number of $(2+1)$-d Majorana fermions propagating on the surface. To understand the topological classification in more generic superconductors without relying on the mean-field theory, topological field theory approaches have been proposed, including gravitational topological response theories\cite{ryu2012electromagnetic,wang2011topological} and a ${\rm U(1)}$ axion field theory describing the electromagnetic response\cite{qi2013axion}. 

Although the free fermion classification of topological states is stable against interaction in many examples such as integer quantum hall states, this does not hold in general. The first example of SPT with classification qualitatively modified by electron interaction was given by Ref. \cite{fidkowski2010effects}, where the 1D Majorana chain with a properly defined time-reversal symmetry is shown to have an integer classification without interaction and have a $\mathbb{Z}_8$ classification with interaction. In this example, the result can be understood by explicitly studying the interaction effect of the (zero-dimensional) edge states. In higher dimensions, it is difficult to study the interaction effect explicitly.
Therefore, new techniques have been developed, including the analysis of the topological surface states\cite{fidkowski2013non,wang2014interacting,metlitski2014interaction,Kitaev} and topological field theory approaches\cite{kapustin2014fermionic,you2014symmetry,EdWitten}. 
It is predicted that the integer classification of DIII class TSC in $(3+1)$-dimensions by the topological band theory reduces to $\mathbb{Z}_{16}$ when interaction is considered. In other words, TSCs with topological quantum number $\nu=1,2,...,15$ are robust against interaction, while $\nu=16$ becomes trivial with interaction.  

In this paper, we provide an alternative understanding of the $\mathbb{Z}_{16}$ classification of interacting TSCs by studying anomalies in electromagnetic responses. More specifically, we studied the vortex crossing process in type-II topological superconductors. We show that an anomalous Berry's phase is obtained in the event of a crossing between ``chiral vortex line" (which will be defined later in the text) and an ordinary superconducting vortex line. By relating the $(3+1)$-d TSC to a $(4+1)$-d TI, this anomaly can be understood as a charge pumping in the extra dimension. This Berry's phase vanishes for a $\nu=16$ topological superconductor, consistent with the previous prediction that the topological classification is $\mathbb{Z}_{16}$. Our result can be obtained by applying the axion field theory proposed in Ref. \cite{qi2013axion}, but can also be understood based on more general arguments without referring to the low energy effective field theory. We generalize our discussion of DIII TSC to other $d+1=4n$ space-time dimensions. The quantum anomaly predicts a $\mathbb{Z}_\nu$ classification, with $\nu=2^{(d+5)/2}$ in $8k+4$ space-time dimensions (with $k$ an integer), and $\nu=2^{(d+3)/2}$ in $8k$ space-time dimensions. We compare our results with other approaches, including the non-linear $\sigma$ model approach\cite{Kitaev, you2014symmetry} and the $\eta$ invariant approach\cite{EdWitten}, and find consistency.  



\section{A brief review of the axion field theory for $(3+1)$-d TSC}

In this section we will review the axion field theory for $(3+1)$-d TSC developed in Ref. \cite{qi2013axion}. Electromagnetic field is suppressed in a superconductor due to the Anderson-Higgs mechanism. However, in type-II superconductors magnetic field can penetrate into the superconductor through vortices with quantized flux $hc/2e$ (equals $\pi$ in the natural units). This enables the possibility of defining topological effects in electromagnetic response, which will be our main interest.


\subsection{Derivation of the axion field theory in $(3+1)d$}

Let's first recap the strategy in the paper\citep{qi2013axion}: in order to derive an effective field theory for topological response perturbatively, we are going to consider an insulating state with ${\rm U(1)}$ symmetry as auxiliary system. For instance, in the $(3+1)$-d axion field theory for topological superconductors, the auxiliary insulating state in the derivation is a $(4+1)$-d topological insulator\cite{qi2008topological} which (without interaction) has an integer classification. The integer topological invariant $c_2$ is the second Chern number of the band structure, and corresponds to the net helicity of $(3+1)$-d Weyl fermions on its boundary. The minimal surface theory of a system with $c_2=n$ consists of $n$ left-hand Weyl fermions on one boundary and $n$ right-hand Weyl fermions on the opposite boundary. The electromagnetic response of this system is described by the $(4+1)$-dimensional Chern-Simons term:

\begin{eqnarray}
S_{CS}[A]&=&\frac{c_2}{24\pi^2} \int d^5 x \epsilon^{abcde} A_a \partial_b A_c \partial_d A_e\label{eq:CS5}
\end{eqnarray}


\begin{figure}[htb]
\centering
\includegraphics[scale=0.4]{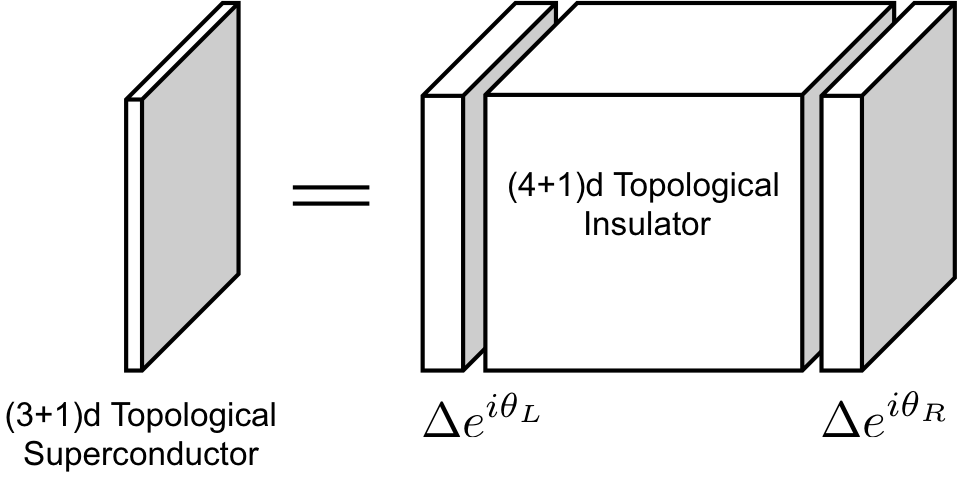}
\caption{A convenient model of $(3+1)d$ superconductor
on the spatial manifold $M_3$ is given by a
$(4+1)d$ time-reversal invariant topological insulator defined on $M_3 \times I $, in proximity with two s-wave superconductors with the order parameter $\Delta e^{i\theta_L}$ and $\Delta e^{i\theta_R}$. When $\theta_L-\theta_R=0$ ($\pi$) the superconductor is trivial (topological), respectively. }
\label{tsc}
\end{figure}

%
%
 As is illustrated in Fig.\ref{tsc}, the $(3+1)$-d topological superconductor on the spatial manifold $M_3$ can be represented as the boundary state of a $(4+1)$-d topological insulator defined on a slab $M_3 \times I $, coupled to two $s$-wave superconductors on each side with order parameters $\Delta e^{i\theta_L}$ and $\Delta e^{i\theta_R}$. Through a gauge transformation, the phase difference $\theta_L-\theta_R$ of the two surface superconductors is equivalent to a vector potential $A_w$ in the extra spatial direction $w$. This vector potential is coupled to the electromagnetic field in the physical $(3+1)$-d by the Chern-Simons term(\ref{eq:CS5}), leading to the dimensionally reduced action of the $(3+1)$-d theory defined on $M_3$:
\begin{eqnarray}
S_{top.}[\theta,A]&=&\frac{c_2}{64\pi^2} \int d^4 x \epsilon^{\mu\nu\sigma\tau} \theta  F_{\mu\nu} F_{\sigma\tau}\label{action} \\
\theta &=& \theta_L-\theta_R\nonumber
\end{eqnarray}
The topological invariant $c_2$ of the $(4+1)$-d theory is related to that of the $(3+1)$-d topological superconductor. In a time-reversal invariant superconductor $\theta_L,\theta_R=0,\pi$, and the topological invariant of TSC is given by
\begin{eqnarray}
\nu=\frac{c_2}2\left(e^{i\theta_L}-e^{i\theta_R}\right)
\end{eqnarray}
Without losing generality, we will assume we are working with a TSC whose $(\theta_L,\theta_R)=(0,\pi)$, such that $\nu=c_2$.

The full action should also includes the Higgs term, Maxwell term and Josephson coupling between $\theta_L$ and $\theta_R$. The long-wavelength behavior of electromagnetic field is dominated by the Higgs term, but the topological term (\ref{action}) remains significant in non-perturbative processes of vortex crossing. The Josephson coupling between $\theta_L$ and $\theta_R$ is allowed by symmetry of the system, and leads to confinement of vortices of $\theta_L$ with that of $\theta_R$ in the bulk of the superconductor. 

\subsection{Topological response: Fermion parity pumping}

\begin{table}[hbt]
\centering
\begin{tabular}{c|c|c|c}
& lefthanded & righthanded & non-chiral\\
 \hline
$\theta_L$ & $2\pi$& 0 & $2\pi$ \\
 \hline
$\theta_R$ & $0$& $2\pi$ & $2\pi$ \\
 \hline
EM & $\pi/2$& $\pi/2$ & $\pi$
\end{tabular}
\caption{Table for string-like excitations in a TSC. The lefthanded and righthanded chiral vortices both contain $\pi/2$ EM flux, while the combination of them form the non-chiral vortex, {\it i.e.} the usual superconducting vortex, carrying flux $\pi$.}
\label{table}
\end{table}

\begin{figure}[htp]
\centering
\includegraphics[scale=0.3]{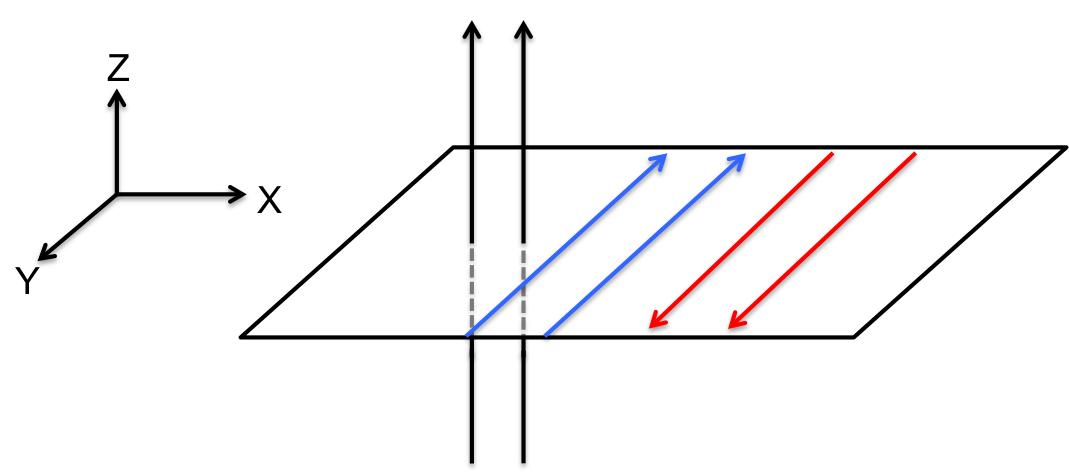}
\caption{A configuration of lefthanded (blue) and righthanded (red) chiral vortices, and non-chiral vortices (black) in a system defined on the three-dimensional torus $T^3$ (see text). }
\label{periodic}
\end{figure}

The axion field theory only depends on three fields: $\theta_L$, $\theta_R$ and electromagnetic field $A_\mu$, so the elementary excitation are three types of vortex strings: left hand chiral vortex string of $\theta_L$, right hand chiral vortex string of $\theta_R$ and their combination, which we will call a non-chiral vortex string, see Table(\ref{table}). Due to flux quantization of the electromagnetic gauge field, there is always even number of non-chiral vortex strings crossing any closed surface. In particular if we consider a system with the spatial geometry of 3 dimensional torus $T^3$, and consider straight vortex lines in $y$ and $z$ directions as is shown in Fig.\ref{periodic}, there must be at least two non-chiral vortex strings, or two left-hand chiral vortices and two right-hand chiral vortices. In a periodic space-time process, each vortex string must return to its original position, so that $8$ crossings occur between these vortices if we move the vertical vortices across the horizontal ones. 




Due to the Higgs term, the electromagnetic field concentrates in a small neighborhood of the vortex strings (in the scale of penetration depth), so that the topological Lagrangian density $\epsilon^{\mu\nu\sigma\tau} \theta  F_{\mu\nu} F_{\sigma\tau}$ is always zero except at the crossing point of two strings. In the space-time periodic process with $8$ vortex crossings, the net effect is a minimal quantum of $\int d^4x \epsilon^{\mu\nu\sigma\tau}F_{\mu\nu} F_{\sigma\tau}=32\pi^2$, and the axion term describes a pumping of a single fermion from left-hand fermions to right-hand fermions. Since the Cooper pair has the charge of two electrons, such a pumping of a single fermion is still well-defined despite of the Cooper pair condensation. In other words, what is changed during this vortex crossing process is the fermion number parity of the fermions with each chirality.


\subsection{Confinement of chiral vortices}

\begin{figure}[htb]
\includegraphics[scale=0.3]{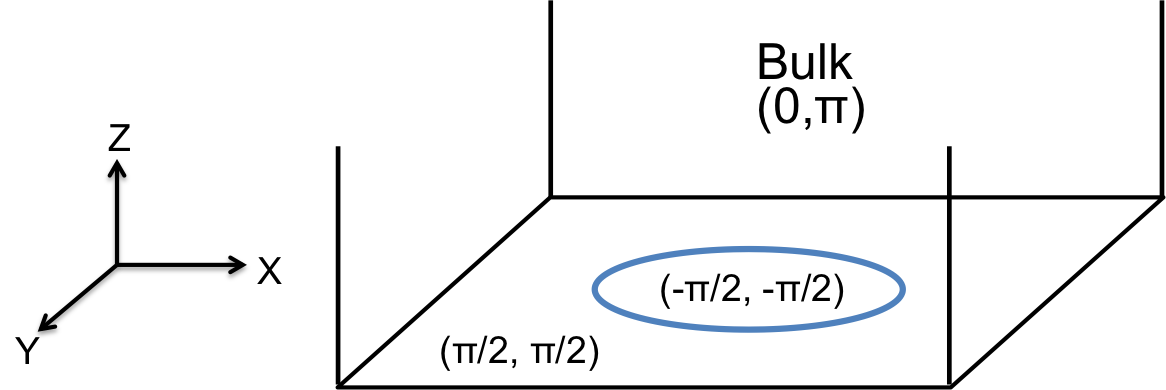}
\caption{An example of chiral string on surface, fixed by boundary conditions: e.g. $(\theta_L,\theta_R)$ in the bulk is $(0,\pi)$; on the surface, two T breaking region carries $(-\pi/2,-\pi/2)$ and $(\pi/2,\pi/2)$ separately.}\label{fig:surfacestring}
\end{figure}

In the discussion above we have focused on topological properties of vortices. When we consider the energetics of vortices, it should be noticed that the chiral vortices with opposite chirality are actually confined in the bulk of the superconductor, because there is no separate ${\rm U(1)}$ symmetry of left-hand and right-hand order parameters $\theta_L$ and $\theta_R$. In general the action of the theory contains a Josephson term
\begin{eqnarray}
J\cos (\theta_L-\theta_R)
\end{eqnarray}
which physically comes from the inevitable coupling of left-hand and right-hand fermions at high energy. This term leads to a linear confinement energy between vortices with opposite chirality. However, chiral vortex strings can still survive on the interface between TSC and a trivial superconductor, which correspond to a domain wall between two boundary conditions at the interface, as is shown in Fig.\ref{fig:surfacestring}. The two domains on the surface are related by time-reversal symmetry, so that they have identical energy. Consequently, the chiral vortices on the surface are not confined.

One important consequence of this confinement is that the vortex crossing process can not happen between two chiral strings, because two chiral strings always stay in the same plane. Hence, the ``minimal" vortex crossing event is that between a chiral string on the surface and a non-chiral string perpendicular to the surface.

\section{$\mathbb{Z}_{16}$ classification}

\subsection{Fractional charge pumping}\label{sec:chargepumping}

\begin{figure}[htb]
\includegraphics[scale=0.3]{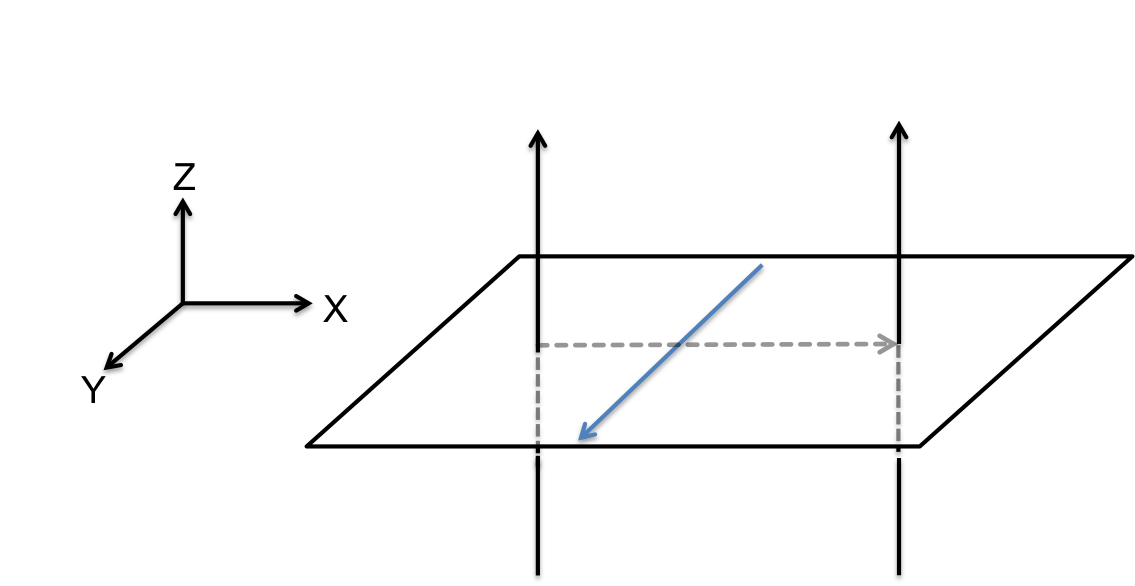}
\caption{A single crossing event between chiral(blue) and non-chiral(black) vortex strings.}\label{fig:singlecross}
\end{figure}

In the previous section, we recalled some results in Ref.\onlinecite{qi2013axion} that are relevant to the current work. In this section, we will obtain a new result, the $\mathbb{Z}_{16}$ classification of TSC, by studying a single/primitive vortex crossing event, as is shown in Fig.\ref{fig:singlecross}. Such an event involves one chiral vortex string and one non-chiral vortex string, which corresponds to an electromagnetic field strength of
\begin{eqnarray}
F^{\text{chiral}}_{zx}&=&\frac{\pi}{2} \delta(x-x_0) \delta(z-z_0) \\
F^{\text{non-chiral}}_{xy}&=&\pi \delta(x-vt-x_0)\delta(y-y_0) \\
F^{\text{non-chiral}}_{ty}&=&-v\pi \delta(x-vt-x_0)\delta(y-y_0)
\end{eqnarray}
with all other components zero, except those related to above by symmetry of indices. (Note that the non-chiral string is moving, so it induces an electric field along $y$ direction $F_{ty}$.)

\begin{figure}[htb]
\includegraphics[scale=0.4]{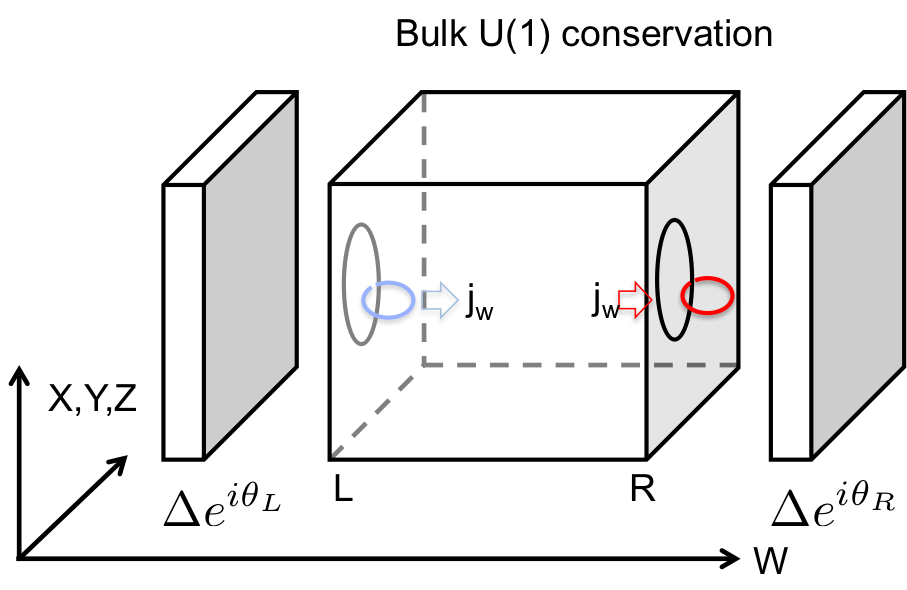}
\caption{Charge pumping through the extra dimension.}
\end{figure}


To understand the consequence of a single vortex crossing, it is most intuitive to use the relation of TSC to $(4+1)$-d topological insulator that was discussed in Fig.\ref{tsc}. In the following we will consider a TSC obtained from a slab of $(4+1)$-d TI with superconductors on the surface and explain the intuitive reason of $\mathbb{Z}_{16}$ classification. Then we provide more rigorous arguments to show that the result applies to generic TSC and is independent from this particular model.

During the single vortex crossing event, a charge current is introduced in the bulk, described by the Chern-Simons term (we are considering $c_2=\nu=1$):
\begin{eqnarray}
j_w = \frac{\delta S_{CS}}{\delta A} 
= \frac{1}{32\pi^2} \epsilon^{\mu\nu\sigma\tau} F_{\mu\nu} F_{\sigma \tau}
\end{eqnarray}

Since $F$ is only non-vanishing in the boundary $(3+1)d$ system, the only nonzero component of the current is in the direction perpendicular to the physical $(3+1)$-dimensions, which we denote as $w$ component. The net charge pumped from the left-hand superconductor to the right-hand one is obtained by integration over $j_w$, which is 
\begin{eqnarray}
\Delta Q_w =\frac{1}{32\pi^2} \times 2\times 2^2 \times \pi/2 \times \pi  
=\frac{1}{8}
\end{eqnarray}
during a single vortex crossing event. 
This fractional charge inflow is analogous to the charge accumulation in integer quantum hall effect induced by a flux threading\cite{weeks2007anyons}, which is a consequence of the $(2+1)$-d Chern-Simons term. 
In the $(3+1)$-d point of view, a fractional charge $1/8$ is pumped from the left-handed fermion to the right-handed fermion, so that no net charge flow has occured, and what happens is a chiral charge pumping.


The discussion above applies to a minimal TSC with topological invariant $\nu=1$. For a TSC with topological invariant $\nu$ the charge pumped during a single vortex crossing is $\frac{\nu}8$. Since charge is only well-defined modulo $2$ in a superconductor, the charge pumping effect becomes trivial for $\nu=16$. This is why the topological classification of TSC is reduced to $\mathbb{Z}_{16}$.


\subsection{An alternative explanation: chiral anomaly}\label{sec:anomaly}

\begin{figure}[htb]
\centering
\includegraphics[scale=0.35]{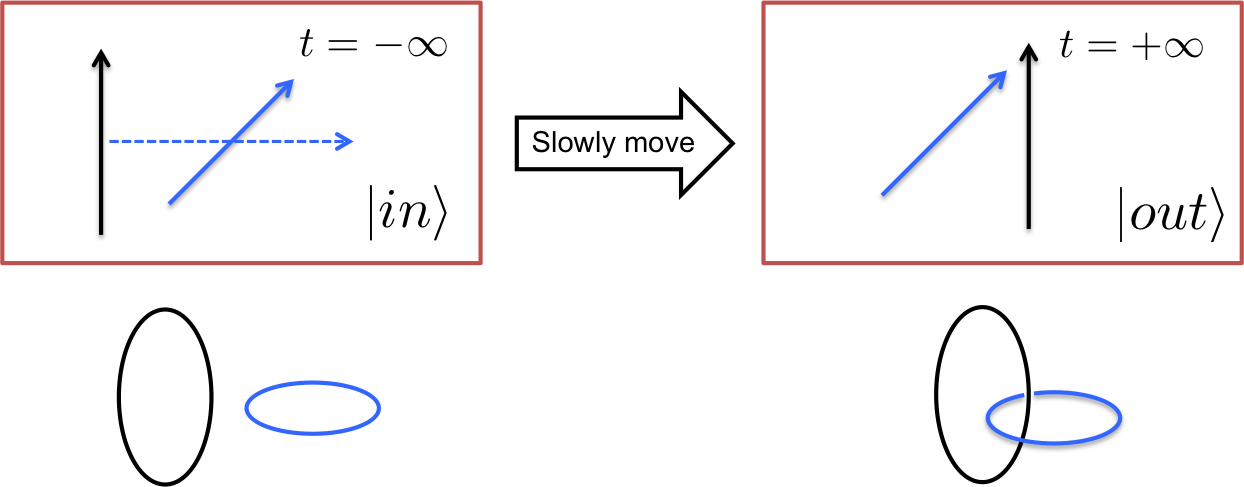}
\caption{$| \rm in \rangle $ and $ | \rm out \rangle $ states associated to a vortex crossing event.}
\label{single}
\end{figure}

In the previous subsection we have described the consequence of vortex crossing as a charge pumping in the $(4+1)$-d picture. To understand whether this argument applies to generic $(3+1)$-d TSC we would like to reformulate the argument in a way that is applicable without referring to the $(4+1)$-d theory. For this purpose, we investigate the consequence of chiral anomaly in TSC.
A chiral ${\rm U(1)}$ rotation is defined by shifting the superconducting phase $\theta_L$ without shifting $\theta_R$. Even at presence of the Josephson coupling term $J\cos\left(\theta_L-\theta_R\right)$, the shift of $\theta_L$ by $2\pi$ should still be a symmetry:
\begin{eqnarray}
s_{L}: \theta_{L} &\rightarrow & \theta_{L}+2\pi
\end{eqnarray}
Alternatively, one can also define the shift of $\theta_R$ without shifting $\theta_L$. Since $s_L$ is expected to be a global symmetry for this system, no physical process should depend on this transformation. However, during the single vortex crossing event, even when the movement is infinitely slow, the system still obtains a nontrivial Berry's phase due to the topological term:
\begin{eqnarray}
\langle {\rm out} | U(\infty,-\infty) | {\rm in} \rangle &=& e^{iS_{top}}=\exp\left[i\frac{\nu}{16}\left(\theta_L(x_0)-\theta_R(x_0)\right)\right]\nonumber\\\label{adiabatic}
\end{eqnarray}
Here $U\left(\infty,-\infty\right)$ is the adiabatic time-evolution operator, and states $|{\rm in}\rangle$ and $|{\rm out}\rangle$ are the initial and final states before and after the vortex crossing event, respectively (see Fig.\ref{single}).
The fact that this action is not invariant in the symmetry transformation $s_L$ is known as the chiral anomaly. 

Compare the anomaly discussion with the $(4+1)$-d picture, the phase obtained by the system in transformation $\theta_L\rightarrow \theta_L+2\pi$ is $e^{i\pi Q_L}$, which is measuring the charge of the left-hand fermion (mod $2$). Therefore the fact that the phase changes by a factor of $e^{i\pi \nu/8}$ is equivalent to the charge pumping from one surface of the $(4+1)$-d TI to the other surface that we discussed in the previous subsection.

An obvious concern about equation (\ref{adiabatic}) is that the Berry's phase for a single vortex crossing process is not gauge invariant, since the system does not return to the same state. By changing the definition of state $|{\rm in}\rangle$ and $|{\rm out}\rangle$ one can shift the phase $\left\langle {\rm out}\right|U(\infty,-\infty)\left|{\rm in}\right\rangle$ by an arbitrary phase factor. However, the anomalous dependence of this phase factor to $\theta_L$ and $\theta_R$ is independent from such gauge choice and is a well-defined quantity. To be more precise, we can denote
\begin{eqnarray}
\left\langle {\rm out}\right|U(\infty,-\infty)\left|{\rm in}\right\rangle=e^{i\varphi (\theta)}\nonumber\\
\Delta \varphi \equiv \int_0^{2\pi}d\theta \frac{\partial \varphi}{\partial \theta}=\frac{\pi}8\nu\label{Deltaphi}
\end{eqnarray}
with $\theta=\theta_L-\theta_R$. The change of $\theta$ in an adiabatic change of $\theta_L-\theta_R$ by $2\pi$ is gauge invariant modulo $2\pi$, which defines a physical procedure of measuring $\nu$ modulo $16$.

From the discussion in this subsection, we see that the fractional charge pumping is equivalent to a Berry's phase effect obtained by tuning two parameters, vortex crossing and the value of $\theta_L-\theta_R$ at the crossing point. This reformulation does not explicitly depend on the $(4+1)$-d picture, which thus confirms that our results on the $\mathbb{Z}_{16}$ classification is applicable to generic $(3+1)$-d TSC. Furthermore, the Berry's phase formulation also indicates that it is possible to provide an even more general classification of $(3+1)$-d TSC which does not even need to rely on the topological field theory description. This will be the task of next section.

\section{General topological argument}
\label{general}

In the previous section, we offers an argument on why we need 16 copies of $\nu=1$ TSC to trivialize the physical consequence of vortex line crossing, which is based on the topological action. In this section, we would like to develop a more general approach to the $\mathbb{Z}_{16}$ classification based on general geometrical arguments and general properties of superconductors, without explicitly using the topological action. 

\begin{figure}[h]
\centering
\includegraphics[scale=0.2]{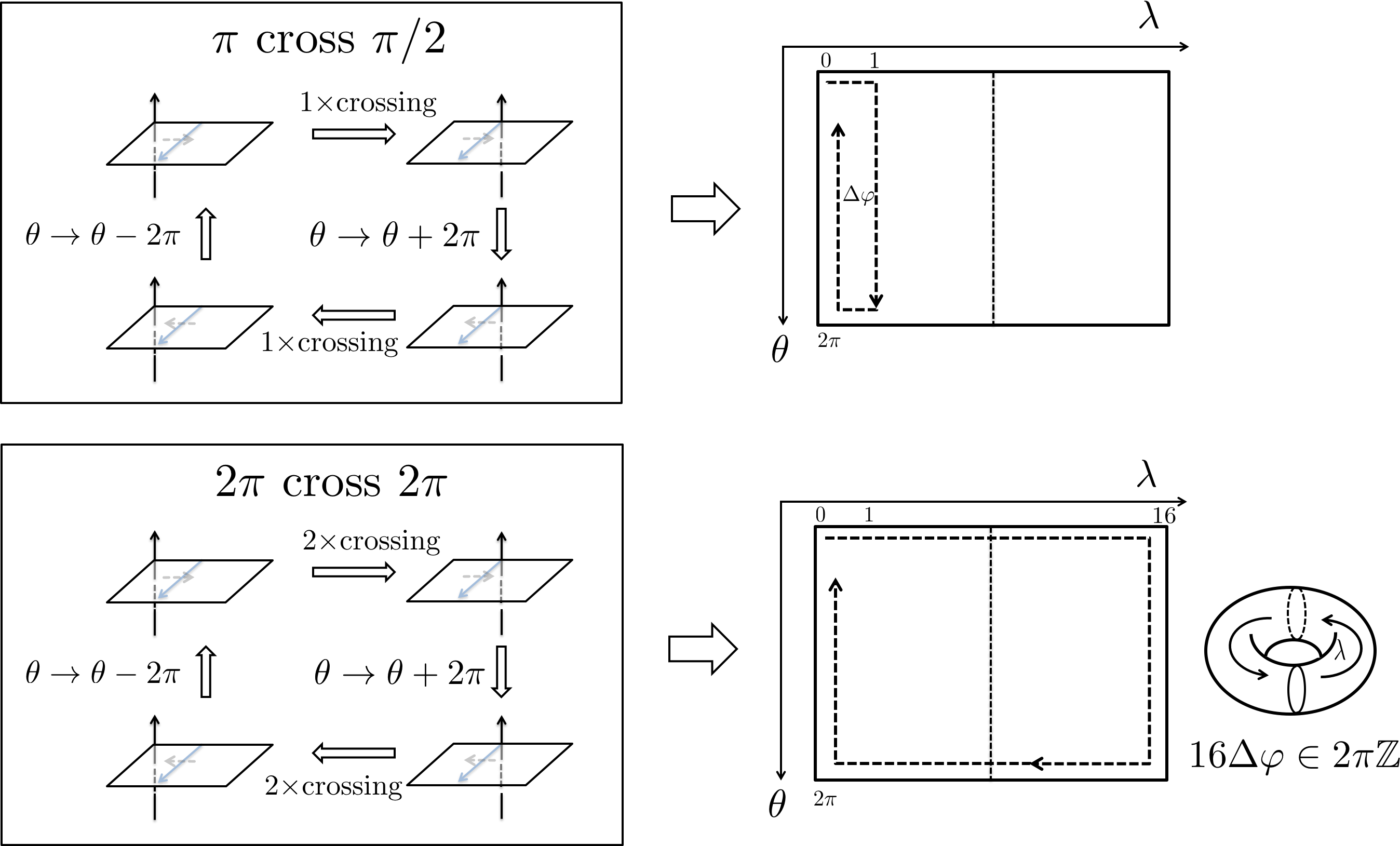}
\caption{Illustration for the Berry's phase $\Delta \varphi$ defined for the parameter space $(\theta, \lambda)$. The $\theta$ direction is periodic with period $2\pi$. The Berry's phase obtained by adiabatic variation of $\theta$ from $0$ to $2\pi$ varies with $\lambda$, determined by the Berry curvature. Once a period in $\lambda$ is found (see text), a torus is defined in the $(\theta,\lambda)$ space, and the flux ({\it i.e.} first Chern number) of the Berry curvature in this torus defines the topological invariant (see text).  
}
\label{fig:general}
\end{figure}

We start by defining a two-dimensional parameter space for a superconductor with vortices (Fig.\ref{fig:general}), which is a reformulation of the discussion in Sec.\ref{sec:anomaly} in more rigorous language. One parameter is the chiral $\theta$ angle of the material, which should be considered as a parameter that adiabatically interpolates between time-reversal invariant trivial superconductor at $\theta=0$ and TSC at $\theta=\pi$. For other values of $\theta$ the system breaks time-reversal symmetry. In the case of a superconductor with weak pairing amplitude near the Fermi surface, $\theta$ reduces to $\theta_L-\theta_R$, the difference of order parameter phases on the two types of Fermi surfaces, but in general $\theta$ can be more general interpolation parameter between trivial and topological superconductors. The detail of the choice of $\theta$ does not matter. The only property important here is that there exists an adiabatic path to connect trivial superconductor and TSC as long as we are allowed to break time-reversal symmetry. In addition we choose $\theta$ such that $\theta\rightarrow -\theta$ upon time-reversal transformation. The second parameter, denoted as $\lambda$, describes the position of a vortex string. Tuning $\lambda$ from $0$ to $1$ leads to a vortex crossing between a chiral vortex string (i.e. a vortex of $\theta$) and a non-chiral vortex string.

The ground state of the system $\left|G(\theta,\lambda)\right\rangle$ varies with these two parameters, which defines a Berry connection $a_s=-i\langle G|\partial_s|G\rangle$, with $s=\theta,\lambda$. The Berry curvature is
\begin{eqnarray}
F_{\theta\lambda}=-i\left[\frac{\partial\langle G|}{\partial \theta}\frac{\partial |G\rangle}{\partial \lambda}-\frac{\partial\langle G|}{\partial \lambda}\frac{\partial |G\rangle}{\partial \theta}\right]
\end{eqnarray}
When we consider the vortex crossing at a given background value of $\theta$, the phase obtained is
\begin{eqnarray}
\varphi(\theta)=\int_0^1 d\lambda a_\lambda(\theta,\lambda)
\end{eqnarray}
which is not gauge invariant. The winding of $\varphi(\theta)$ during change of $\theta$ by $2\pi$ is however gauge invariant, which is determined by the surface integration of gauge curvature:
\begin{eqnarray}
\Delta \varphi=\int_0^{2\pi} d\theta \frac{\partial \varphi}{\partial \theta}=\int_0^{2\pi} d\theta\int_0^1 d\lambda F_{\theta\lambda}
\end{eqnarray}
$\Delta\varphi$ is a physical property of a given parameterized family of superconductors. Without further contraints to it, it is not obvious how to use $\Delta\varphi$ to probe the discrete topological invariant in the TSC. In Eq.(\ref{Deltaphi}), we used the knowledge from topological action to show that $\Delta\varphi$ is quantized to $\nu \frac{\pi}8$. In the following we would like to provide a more general geometric argument why $\Delta \varphi$ is quantized even if we don't rely on the topological action.

To understand the quantization of $\Delta\varphi$, the key is to construct a vortex-crossing process that is periodic. If we label the vortex configuration during a periodic process also by $\lambda$, the parameter space $(\theta,\lambda)$ is now a torus since it is periodic in both directions. Therefore the net flux $\int_{T^2} d\theta d\lambda F_{\theta\lambda} \in 2\pi \mathbb{Z} $ is quantized in unit of $2\pi$, which leads to the quantization of $\Delta \varphi$, see Fig(\ref{fig:general}).

In Ref.\cite{qi2013axion}, a crossing between four chiral vortices and two non-chiral vortices are considered in a torus $T^3$. Because the net electromagnetic flux in each $2$-cycle of the torus is quantized in the unit of $2\pi$, the number of superconducting vortices, each with flux $\pi$, penetrating each cycle must be even. Since each superconducting vortex corresponds to two chiral vortices, a single crossing between two $2\pi$ flux tubes in $T^3$ corresponds to $8$ crossings between $4$ chiral vortices and $2$ non-chiral ones. Therefore naively one would conclude that $8$ vortex crossings is a periodic process. However, $8$ vortex crossings may actually still have a nontrivial effect to fermions, which can be seen from the following argument.

\begin{figure}[htb]
\centering
\includegraphics[scale=0.2]{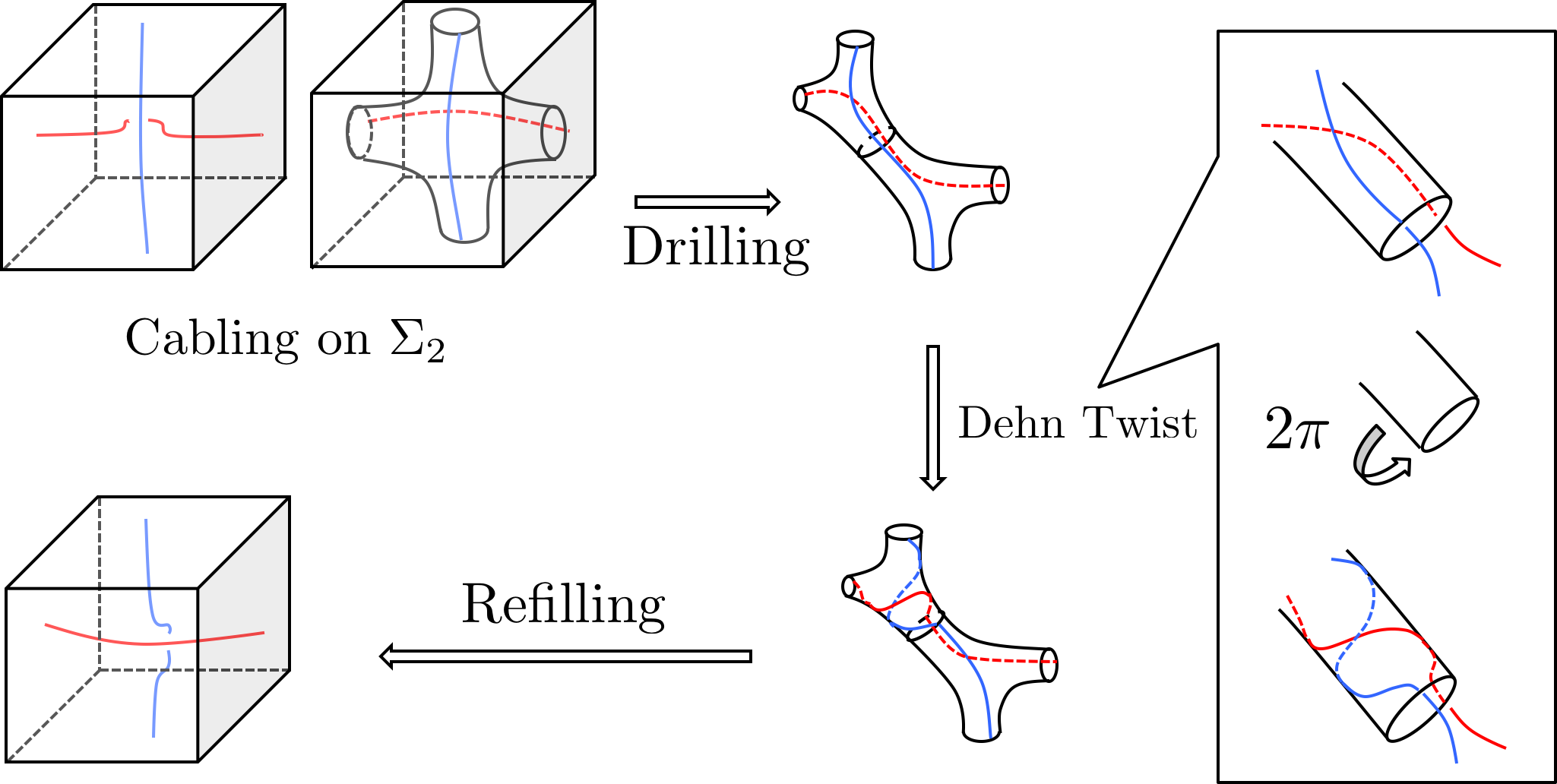}
\caption{Illustration of the fact that the crossing between a pair of $2\pi$ vortices in $T^3$ is equivalent to a surgery with a Dehn twist. Here, we represent $T^3$ by a cube with opposite face identified. }
\label{dehn}
\end{figure}

As shown in Fig.\ref{dehn}, we consider two $2\pi$ vortices winding around two different cycles in $T^3$. 
The crossing of these two vortices is equivalent to a surgery procedure: Consider a genus-$2$ Riemann surface $\Sigma_2$ (Fig.\ref{dehn}) covering both vortex lines. The surgery is defined by cutting the $T^3$ at the surface $\Sigma_2$ into two domains, and gluing them back after a Dehn twist along a curve in $\Sigma_2$ (black circle in the middle of $\Sigma_2$ in Fig.\ref{dehn}).


More explicitly, the procedure in the Fig.\ref{dehn} includes 3 steps: (1)``Cable" the vortex strings in $T^3$ right beneath a genus $2$ surface $\Sigma_2$ to ensure it belongs to the interior; (2) Drill the genus $2$ handlebody(interior) out, and implement a Dehn twist on its surface; (3) Refill the handlebody with the Dehn twist, which leads to a closed 3 manifold. From Fig.\ref{dehn} one can see that the two vortex lines are winded around each other during the Dehn twist, such that their final position is equivalent to that after a vortex crossing. Using Heegaard diagram (see Appendix(\ref{proof}) for technical proof.), we can prove that the surgery described above does not change the topology of the hosting 3-manifold $T^3$. Therefore, after moving the flux through the periodic boundary, the manifold with two strings returns to the starting configuration. Through this topological argument, we represent a crossing between two $2\pi$ flux tubes by a surgery with a Dehn twist on the surface $\Sigma_2$. (There is a subtlety if we consider the flux loops as framed ({\it i.e.} ribbons) instead of featureless lines, which we 
will consider in Appendix(\ref{final}).)

Remarkably, although the Dehn twist we consider preserves the topological structure of the manifold, it in general will change the \textit{spin structure} of the surface.\footnote{It is possible that Dehn twist keeps some specially chosen spin structure invariant. For instance, a torus with periodic boundary conditions on both directions is invariant under modular group. On the contrary, the other three spin structures are related by modular group actions.} A system with fermions is sensitive to the spin structure, since the fermion boundary conditions are different for different choices of spin structure. To be more precise, we can consider doing the surgery physically by cutting the TSC at surface $\Sigma_2$. A massless Majorana fermion surface state will be induced on $\Sigma_2$ if time-reversal symmetry is preserved. Therefore the boundary condition of this surface state is sensitive to the spin structure of $\Sigma_2$. Only if we consider a Dehn twist procedure that preserves the spin structure of $\Sigma_2$, we can be sure that the surface states resulting from the surgery still have the same boundary condition after the Dehn twist, and therefore can be glued again to a gapped TSC.

Due to this consideration, we see that a single crossing between two $2\pi$ vortices may not be considered as a periodic process. 
However, it can be shown that carrying the same Dehn twist twice always preserves the spin structure of the surface (See Appendix(\ref{proof}) for technical details and proof). Consequently, we conclude that two crossings between a pair of $2\pi$ vortices is always a periodic process, which leads to the same vortex configuration in $T^3$ with the same spin structure. Note that the $2$ crossings of $2\pi$ vortices are equivalent to $16$ crossings of chiral vortices with non-chiral vortices, we conclude that the phase factor $\Delta\varphi$ obtained in $16$ crossings must be quantized in the unit of $2\pi$. Therefore $\Delta\varphi$ for each crossing satisfies
\begin{eqnarray}
\Delta\varphi=\nu \times \frac{2\pi}{16},~\nu\in\mathbb{Z}
\end{eqnarray}

In summary, by mapping a vortex crossing process to a surgery on $T^3$ with a Dehn twist involved, we learn that the topological invariant $\nu~{\rm mod}~16$ is the first Chern number in the parameter space $(\theta,\lambda)$, where $\theta\in[0,2\pi)$ describes the periodic interpolation between TSC and trivial superconductor, and $\lambda$ describes a periodic vortex crossing process involving $16$ vortex crossing points, which together are equivalent to a double Dehn twist.




\section{Generalization to DIII TSC in $4n$ space-time dimensions}

In this section, we will generalize the axion field theory approach to the topological classification of TSC to higher dimensions. More explicitly, we are going to study the DIII class TSC in space-time dimension $d+1=4n$. We will follow the idea and strategy of the $(3+1)d$ DIII class case, and relate the $4n$-dimensional TSC to the $4n+1$-dimensional time-reversal invariant TI. 

\subsection{Topological insulators in $(4n+1)$ dimensions}

In the same way how $(3+1)$-d time-reversal invariant TSC is related to the $(4+1)$-d time-reversal invariant TI, corresponding to the TRI TSC in $4n$ dimensions there is a $4n+1$-dimensional TRI TI which in the free fermion case is classified by an integer. The electromagnetic response of these TI's are described by Chern-Simons theories\cite{qi2008topological}, with the topological invariant the coefficient of the Chern-Simons term.
\begin{eqnarray}
S_{CS}&=&\frac{c_{2n}}{(2n+1)!(2\pi)^{2n}}\int d^{4n+1}x \epsilon^{\sigma\mu_1\nu_1\mu_2\nu_2...\mu_{2n}\nu_{2n}}\nonumber\\
& & A_\sigma \partial_{\mu_1}A_{\nu_1}\partial_{\mu_2}A_{\nu_2}...\partial_{\mu_{2n}}A_{\nu_{2n}}\label{CS4nplus1}
\end{eqnarray}
In $4n+1$ dimensions, Chern-Simons theories are time-reversal invariant.

Although all $(4n+1)$-d TI's are classified by integer (for non-interacting fermions), there is a subtle difference between $(8k+5)$-d and $(8k+1)$-d ($k \in \mathbb{Z}$), which occurs because of the different quantization of the coefficient $c_{2n}$. For free fermions, $c_{2n}$ is the $2n$-th Chern number of the Berry's gauge field:
\begin{eqnarray}
c_{2n}=\frac{1}{(2n)!(4\pi)^{2n}} \int d^{4n} k \epsilon^{i_1j_1..i_{2n}j_{2n}} \textrm{Tr}[f_{i_1j_1}..f_{i_{2n}j_{2n}}]
\label{cnumber}
\end{eqnarray}
For generic insulators, $c_{2n}$ can take any integer value. However, with the $T^2=-1$ time-reversal symmetry, for even $n$ the Chern number is always an even integer, while for odd $n$ it can be any integer. This statement can be directly verified by computing $c_{2n}$ for lattice Dirac models, i.e., $\hat{H}=\sum_k \hat{\psi}(k)^\dagger  d_j(k) \Gamma^j \hat{\psi}(k)$, with $\Gamma^j,~j=1,2,...,4n$ the Dirac $\Gamma$ matrices satisfying $\left\{\Gamma^j,\Gamma^k\right\}=2\delta^{jk}\mathbb{I}$. 
This conclusion also agrees with the Twisted KR theory computation.\cite{atiyah1966k,freed2013twisted,de2014classificationB}. A more physical way to see the dimension-dependent quantization is to study the minimal surface theory. The surface theories can be written as Dirac fermions preserving charge ${\rm U(1)}$ and time-reversal symmetry. Using the representation of Clifford algebra, one can determine the dimension of the minimal representation, which determines the topological invariant $c_{2n}$. We leave more details of this analysis to Appendix(\ref{higherC}). 
The agreement in fact reflects a deep connection between K-theory and Clifford modules\cite{atiyah1964clifford}.




It is also interesting to comment that the distinction between $\mathbb{Z}$ in $8k+5$ dimensions and $2\mathbb{Z}$ in $8k+1$ dimensions actually have important physical implication on $\mathbb{Z}_2$ topological insulators in adjacient dimensions\cite{qi2008topological}. For $8k+5$ dimension, there are two descendant topological insulators of the same symmetry class achieved by dimensional reduction procedure, which are both classified by $\mathbb{Z}_2$. The most well-known examples are $(3+1)d$ topological insulators and $(2+1)d$ quantum spin hall insulators coming from $(4+1)d$ quantum Hall root states\cite{zhang2001four}. However, for $8k+1$ dimensions, there are no $\mathbb{Z}_2$ descendants, because the doubling of Chern number makes the $\mathbb{Z}_2$ invariants in descendent theories trivial.

\subsection{Dimensional reduction and axion field theory}

\begin{figure}[htb]
\centering
\includegraphics[scale=0.4]{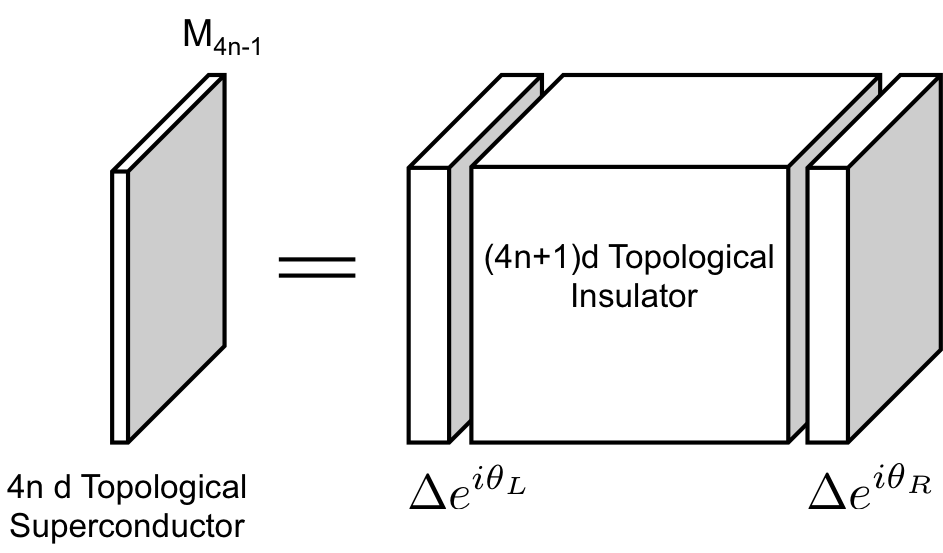}
\caption{An illustration of $4n$ dimensional topological superconductor}\label{fig:hightsc}
\end{figure}

Now we are ready to generalize the axion field theory in $(3+1)$-d to higher dimensions by a very similar dimensional reduction procedure. We consider a $(4n+1)$-d TI defined on a spatial manifold $M_{4n-1} \times I$, with $M_{4n-1}$ a $(4n-1)$-d closed manifold. This system has two boundaries $\partial(M_{4n-1}\times I) =  M_{4n-1}\amalg \overline{M}_{4n-1}$ as is shown in Fig.\ref{fig:hightsc}. On the two surfaces, we introduce proximity coupling to two superconductors with the order parameter $\Delta e^{i\theta_L}$ and $\Delta e^{i\theta_R}$, respectively. The whole system can be viewed as a $4n$-d superconductor, which is a DIII class TRI TSC if $\theta_L,~\theta_R$ take value of $0$ or $\pi$.

The bulk of the $4n+1$-d TI is described by the Chern-Simons theory (\ref{CS4nplus1}). The topological effect of the superconducting phase on the surface can be understood by taking a gauge transformation $A_{4n+1}\rightarrow A_{4n+1}+\partial_{4n+1}\phi,~\theta_L\rightarrow \theta_L+2\phi(0),~\theta_R\rightarrow \theta_R+2\phi(1)$. Here we have labeled the extra-dimensional interval by $x_{4n+1}\in[0,1]$, and label the gauge vector potential along that direction as $A_{4n+1}$. By choosing $\phi\left(x_{4n+1}\right)=-\frac12\left[\theta_L+x_{4n+1}\left(\theta_R-\theta_L\right)\right]$, $\theta_L$ and $\theta_R$ are completely absorbed by $A_{4n+1}=\frac12\left(\theta_L-\theta_R\right)$. After this gauge transformation, we can do the dimensional reduction by assuming all components $A_{\mu}$ are independent from $x_{4n+1}$, which leads to the axion action:
\begin{eqnarray}
S_{top}^{4n}=\frac{c_{2n}}{(2n)!2(4\pi)^{2n}} \int d^{4n}x \ \left(\theta_L-\theta_R\right)F^{2n} 
\end{eqnarray}
with $F^{2n}\equiv\epsilon^{\mu_1\nu_1\mu_2\nu_2...\mu_{2n}\nu_{2n}}F_{\mu_1\nu_1}F_{\mu_2\nu_2}...F_{\mu_n\nu_n}$ as abbreviation. 

\subsection{Vortex crossing and topological classification}

Parallel to $(3+1)d$ case, in general $4n$ dimensions, we can consider the topological response induced by vortex crossing events. A superconducting vortex is always a co-dimension $2$ surface, which means a $4n-2$ dimensional membrane in $4n$-dimensions. To have a topological nontrivial effect, we need to introduce $2n-1$ superconducting vortices and one chiral vortex, which cross at a space-time point. For example, we can consider $2n-1$ superconducting vortices with the field strength
\begin{eqnarray}
F_{2k-1,2k}=\pi\delta(x_{2k-1})\delta(x_{2k}),~k=1,2,...,2n-1
\end{eqnarray}
The intersection of these vortices is a two-dimensional plane along the directions $x_{4n-1},x_{4n}$, which can then cross with a moving chiral vortex that induces a flux $F_{4n-1,4n}=\frac{\pi}2\delta(x_{4n-1})\delta(x_{4n})$. Similar to the $(3+1)$-d case discussed in Sec.(\ref{sec:chargepumping}), the topological response induced by such a crossing is best visualized as a charge pumping in the $(4n+1)$-d TI along the extra direction $x_{4n+1}$:
\begin{eqnarray}
j_{4n+1}
= \frac{c_{2n}}{(2n)!(4\pi)^{2n}} \epsilon^{\mu_1 \nu_1 \ldots \mu_n \nu_n} F_{\mu_1 \nu_1} \ldots F_{\mu_n \nu_n} 
\end{eqnarray}
The net charge pumped during a single vortex crossing event is
\begin{eqnarray}
\Delta Q =\frac{c_{2n}}{(2n)!(4\pi)^{2n}}\times (2n)!2^{2n}\times \frac{\pi}2\times \pi^{2n-1}
= \frac{c_{2n}}{2^{2n+1}}\label{DeltaQhighD}
\end{eqnarray}

Due to the Cooper pair condensation in the surface superconductors, the transfer of charge 2 is considered as trivial. Therefore, the fractional charge $\Delta Q$ mod $2$ is a topological invariant that defines the topological classification of the TSC. In other words, the minimal value of $c_{2n}$ that corresponds to a trivial superconductor is given by $\Delta Q=2$, {\it i.e.} $c_{2n}=2^{2n+2}$. Depending on the space-time dimension, we have the following two cases:
\begin{itemize}
\item For space-time dimension $(d+1)=8k+4$, namely $n=2k+1$, $c_{2n}\in \mathbb{Z}$, so that the TSC is classified by $\mathbb{Z}_\nu$ with $\nu=2^{2n+2}=2^{(d+5)/2}$.
\item
For space-time dimension $(d+1)=8k$, namely $n=2k$, $c_{2n}\in 2\mathbb{Z}$, so that the TSC is classified by $\mathbb{Z}_\nu$ with $\nu=2^{2n+1}=2^{(d+3)/2}$.
\end{itemize}

The charge pumping can be reformulated to an anomaly of the partition function in the chiral rotation of $\theta_L-\theta_R$ in exactly the same way as in the $(3+1)$-d case, which we will not repeat here. 

%
%
%
%
%
%
%

\subsection{Comparison with the $\sigma$-model approach}

Using the axion field theory and vortex crossing, we have obtained a topological classification of DIII class TSC, which is $\mathbb{Z}_\nu$ with
$\nu=2^{(d+5)/2}$ for $d+1=8k+4$, and $\nu=2^{(d+3)/2}$ for $d+1=8k$. We would like to compare this result with other classification schemes. One alternative approach is the non-linear $\sigma$ model analysis\cite{Kitaev}\cite{you2014symmetry}. The surface theory of a $4n$-dimensional TSC is a massless Majorana fermion theory in $(4n-2)$ spatial dimensions, with the Hamiltonian
\begin{eqnarray}
H=  \int d^{4n-2} x\ \psi^T \Gamma^j\left(-i\partial_j\right)  \psi
\end{eqnarray}
Here $\psi$ is a Majorana spinor, and $\Gamma^i$ are real symmetric Dirac matrices. $\Gamma^i$ and the time-reversal matrix $\mathcal{T}$ together form the representation of a real Clifford algebra. Since this Hamiltonian describes a topological surface state, by definition there are no quadratic TRI mass terms allowed. If we consider time-reversal breaking mass terms, it is possible to consider the following massive Hamiltonian:
\begin{eqnarray}
H=  \int d^{4n-2} x \psi^T  \left(-i\Gamma^j \partial_j+\sum_{a=1}^N\gamma^am_a\right)  \psi
\end{eqnarray}
where $\gamma^a$ are purely imaginary anti-symmetric matrices satisfying $\left\{\gamma^a,\gamma^b\right\}=2\mathbb{I}$ and $\gamma^a$ anti-commutes with all $\Gamma^j$. (In other words, $-i\gamma^a$  are negative generators of a real Clifford algebra. )

The mass of the Hamiltonian is $|m|=\sqrt{\sum_a m_a^2}$. The number of mass terms that can be added depends on the number of flavors of Majorana fermions, which is then determined by the topological invariant $\nu$. For example, the minimal surface theory of $(3+1)$-d TSC is $H=\int d^2x \psi^T\left(-i\sigma_x\partial_x-i\sigma_z\partial_y\right)\psi$, with $\sigma_x,\sigma_z$ two Pauli matrices. This Hamiltonian allows only one mass term $\sigma_y$. If we have instead a $\nu=2$ TSC with the surface theory given by $\Gamma_x=\sigma_x\otimes \mathbb{I},~\Gamma_z=\sigma_z\otimes \mathbb{I}$, there can be two mass terms $\sigma_y\otimes \sigma_z,~\sigma_y\otimes \sigma_x$.

The key argument of the $\sigma$ model analaysis is by considering a fluctuating mass $m_a$. By integrating out the massive fermions, one obtains a non-linear $\sigma$ model with the target space $S^{N-1}$. If the $\sigma$ model is in its ordered phase, time-reversal symmetry is spontaneously broken. If the $\sigma$ model is disordered, it is possible to realize a gapped phase with time reversal symmetry restored, which is possible if the bulk is topologically trivial. As was discussed above, the maximal number of mass terms $N$ is determined by the topological invariant $\nu$. If the target space has high enough dimension $N-1>4n$, the $\sigma$ model cannot have any topological term, so that a trivial disordered phase should exist. Following this argument, one obtains a critical value $\nu$ for which the surface state can become trivial due to mass term fluctuation.

Following the idea sketched above, a critical $\nu$ can be determined for all dimensions $4n$ using knowledge about representation theory of Clifford algebra. We will leave more details to Appendix(\ref{fNLSM}), and only state the result that the non-linear $\sigma$ model approach predicts exactly the same classification $\mathbb{Z}_\nu$ as our axion field theory result in the previous subsection. The agreement of these two distinct approaches indicate that the axion theory result is generic.

\section{Conclusion}

In this paper, we revisit the axion field theory description for DIII class topological superconductors in $(3+1)d$. We show that a minimal vortex crossing event between a chiral vortex and a non-chiral vortex leads to a quantum anomaly. We provided two equivalent descriptions of this quantum anomaly. From a purely $(3+1)$-d point of view, the anomaly is a Berry's phase factor obtained during a chiral rotation of the axion phase $\theta_L-\theta_R$. From a $(4+1)$-d point of view, when the TSC is considered as a slab of $(4+1)$-d TI with surface superconductivity, the anomaly corresponds to a charge pumping induced by vortex crossing on the surface. The net charge pumped during a single vortex-crossing event is $\frac{\nu}{8}$. Since in a superconductor, pumping charge $2$ is trivial, this anomaly only detects the topological invariant $\nu$ mod $16$. Therefore our result indicates that the topological classification of $(3+1)$-d TSC is $\mathbb{Z}_{16}$. We also present a generic geometrical argument to show that the anomaly is robustly defined even without referring to the axion topological field theory.

The quantum anomaly argument guarantees that the TSC phases with $\nu=1,2,...,15$ are topologically nontrivial and distinct from each other. However, the absence of anomaly at $\nu=16$ does not necessarily mean $\nu=16$ is trivial. The fact that $\nu=16$ is trivial can be supported by other approaches such as the $\sigma$-model approach we discussed, and the surface topological order approach\cite{fidkowski2013non,wang2014interacting,metlitski2014interaction}.

We further generalize the axion field theory description to arbitrary $d+1=4n$ dimensions to study the topological classification of DIII class TSC in these dimensions. In higher dimensions, the vortex crossing event happens between $2n$ vortices, one chiral and the rest non-chiral. The anomaly suggests that the topological classification is $\mathbb{Z}_\nu$ with the value of $\nu$ increasing exponentially with the spatial dimension. The critical value $\nu$ is $\nu=2^{(d+5)/2}$ for spatial dimension $d=8k+3$, and $\nu=2^{(d+3)/2}$ for $d=8k+7$. As a comparison, we study the topological classification with the non-linear $\sigma$ model approach, which leads to the same conclusion. 

In a recent work\cite{EdWitten}, Edward Witten proposed another topological classification of DIII class TSC by relating it to a different class of topological invariants, the $\eta$ invariants\cite{atiyah1975spectral1}. The topological quantum number of TSC is the coefficient of the $\eta$ invariants in the action. A TSC can be probed by evaluating the partition function on manifolds with different $\eta$ invariants. For the DIII class TSC, the theory can possibly be defined on unorientable manifold with ${\rm pin_+}$ structure. The $\eta$-invariant approach is also applicable to higher dimensions, the result of which, as far as we understand, is consistent with the prediction of axion field theory and $\sigma$ model approach, see footnote\footnote{To be more precise, according to Ref.\cite{gilkey1985eta} (see lemma 2.1 and theorem 3.3 there, which are about $\rm pin_c$, but are also automatically applicable to $\rm pin_+$) the $\eta$ invariant calculation of $\rm pin_+$ manifold $\mathbb{RP}^{4n}$ provides the minimal unit for $\eta$ invariant in dimension $4n$. Moreover, calculation of $\eta$ for $\mathbb{RP}^{4n}$ can be carried explicitly, which predicts the same topological classification as our theory in general $4n$ dimension. We would like to acknowledge helpful discussions with Edward Witten about this point.}. The agreement between these three approaches provides further evidence that the topological classification is robust.

An interesting question is whether our approach to topological classification can be generalized to TI's and TSC's in other dimensions and symmetry classes. It is possible to start from a ``parent state" described by a Chern-Simons theory, and consider spontaneous symmetry breaking. Topological defects in a different symmetry-breaking state may have different co-dimensions as the superconducting vortices, so that a different kind of discrete quantum anomaly may be introduced. The investigation of such generalizations will be left for future works.

\noindent{\bf Acknowledgement}

We would like to acknowledge helpful discussions with Daniel Bulmash, Xie Chen, Chao-Ming Jian, Cenke Xu, Yi-Zhuang You and especially Edward Witten. YG is supported by the National Science Foundation through the grant No. DMR-1151786. XLQ is supported by the David and Lucile Packard Foundation.

\bibliography{refs}

\appendix

\section{Proof of several statements in section \ref{general}}
\label{proof}

In this section, we will provide more detailed and technical proof of two statements in the main text: (A) the surgery in main text(see Fig(\ref{dehn})) with a Dehn twist on the relevant curve does not change the topology of the $T^3$; (B) A double Dehn twist acts trivially on spin structure of surface $\Sigma_g$. In the third subsection, we will mention a subtlety when considering ribbon structure of the strings.

\subsection{Proof of (A)}

To prove statement A, we will recall a useful tool in low dimensional topology called Heegaard diagram. (For reference see Ref.\cite{singer1933three}) We will present the proof after introducing several necessary notions and statements:

Three dimensional closed oriented manifold $M$ can be Heegaard decomposed into two handlebodies $H_1$ and $H_2$ (Warning: the decomposition of $T^3$ in the main text is not a Heegaard decomposition, because the complement is not a handlebody.) $s.t.$ $M=H_1 \cup_f H_2$, where $f$ is a `gluing' map. We denote the `cutting surface' by $\Sigma_g$, which is a genus g Riemann surface. 

On the other hand, given two handlebodies with their boundaries, $(H_1,H_2,\Sigma_g)$, we don't have to know every detail of the gluing map $f$ to reconstruct the manifold. Instead, it is sufficient to use the Heegaard diagram:

\begin{defn}
Heegaard diagram is an ordered pair $(H,\lbrace l_1,\ldots, l_n \rbrace)$, where H is a handlebody, and $\lbrace l_i \rbrace$ is a collection of disjoint, embedded, simple, closed curves on $\partial H$, s.t., $\partial H \backslash \lbrace l_i \rbrace $ is a collection of punched spheres.
\end{defn}

\begin{defn}
Heegaard diagram for $M$: if $\lbrace l_i \rbrace$ are images in a map $f: \partial H_1 \rightarrow \partial H_2$ of boundaries of a system of disks for $H_1$, we say $\lbrace l_i \rbrace$ is a Heegaard diagram for f (or for M).

\textbf{System of disks for H}: a collection of properly embedded, essential disks $\lbrace D_1, \ldots, D_n \rbrace$, s.t. the complement of a regular neighborhood of $\cup D_i$ is a collection of balls.

\textbf{Essential}: Disk $D\subset H$ is essential, if its boundary $\partial D$ doesn't bound a disk on the $\partial H$.
\end{defn}

The main theorem we will use is the following:

\begin{thm}
Let $(H_1,H_2,\Sigma)$ be a Heegaard decomposition of a 3-manifold M and $(H_1',H_2',\Sigma')$ be a Heegaard decomposition of a 3-manifold $M'$, and let $(H_2,\lbrace l_1 \ldots l_n \rbrace)$ be Heegaard diagram for M, $(H_2',\lbrace l'_1 \ldots l'_n \rbrace)$ be Heegaard diagram for $M'$.

If there is a homeomorphism $\phi: H_2 \rightarrow H_2'$ taking each loop $l_i$ to $l_i'$ then there is a homeomorphism $\psi: M \rightarrow M'$.
\end{thm}

An immediate corollary says two identical Heegaard diagram produce the same manifold, upto homeomorphism. Roughly speaking, this theorem indicates that a Heegaard diagram contains adequate information to reconstruct a 3-manifold. This is what we need for proving the statement (A).

\begin{figure}[htb]
\center
\includegraphics[scale=0.2]{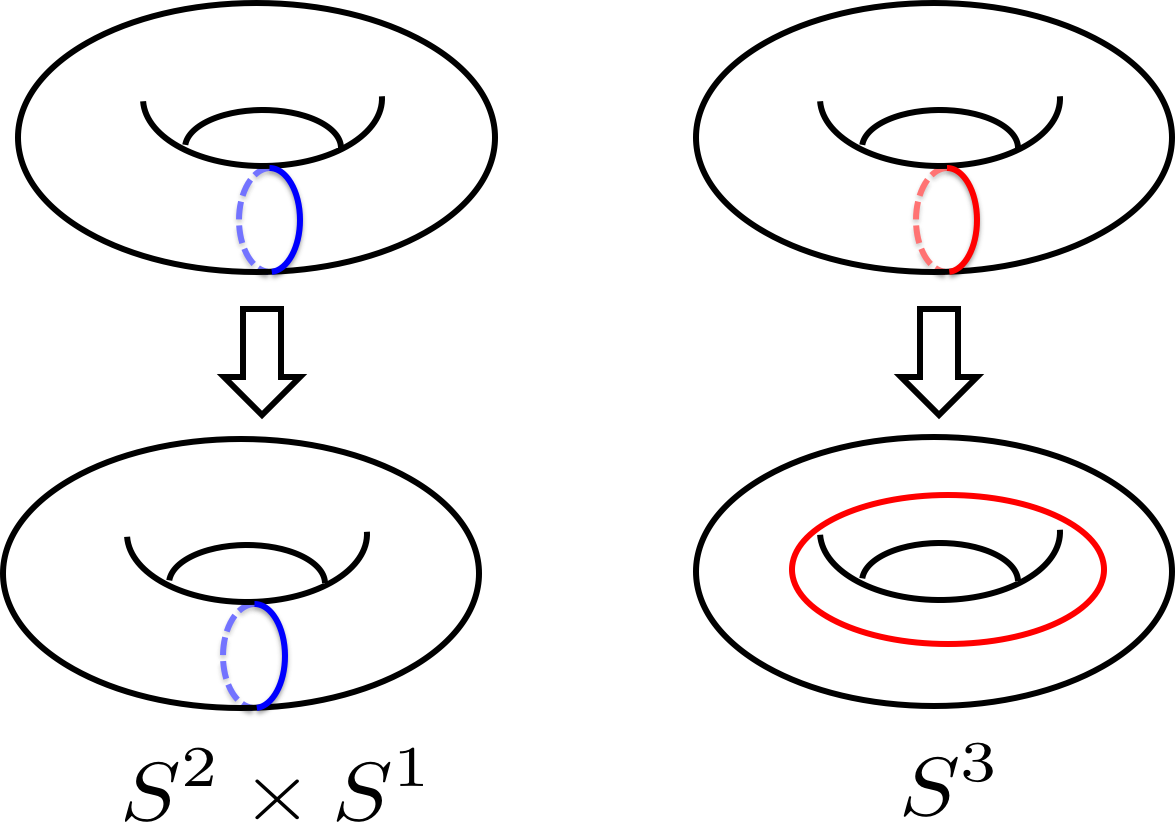}
\caption{Heegaard diagrams for $S^2\times S^1$ and $S^3$.}
\label{example}
\end{figure}

Before proceeding, it is useful to familiarize the tool by a classic example, see Fig.\ref{example}

\begin{ex}
3 sphere $S^3$ v.s. $S^2\times S^1$: 

$S^3$ can be understood as Heegaard decomposition by arbitrary genus $g$ surface. In particular, it could be decomposed by torus with handlebodies being solid torus. 

Similarly, $S^2 \times S^1$ can also be decomposed into two solid torus, but the gluing map is different from the $S^3$ case. The difference can be told from their different Heegaard diagrams, see Fig.\ref{example}.

\end{ex}

\begin{figure}[htb]
\centering
\includegraphics[scale=0.25]{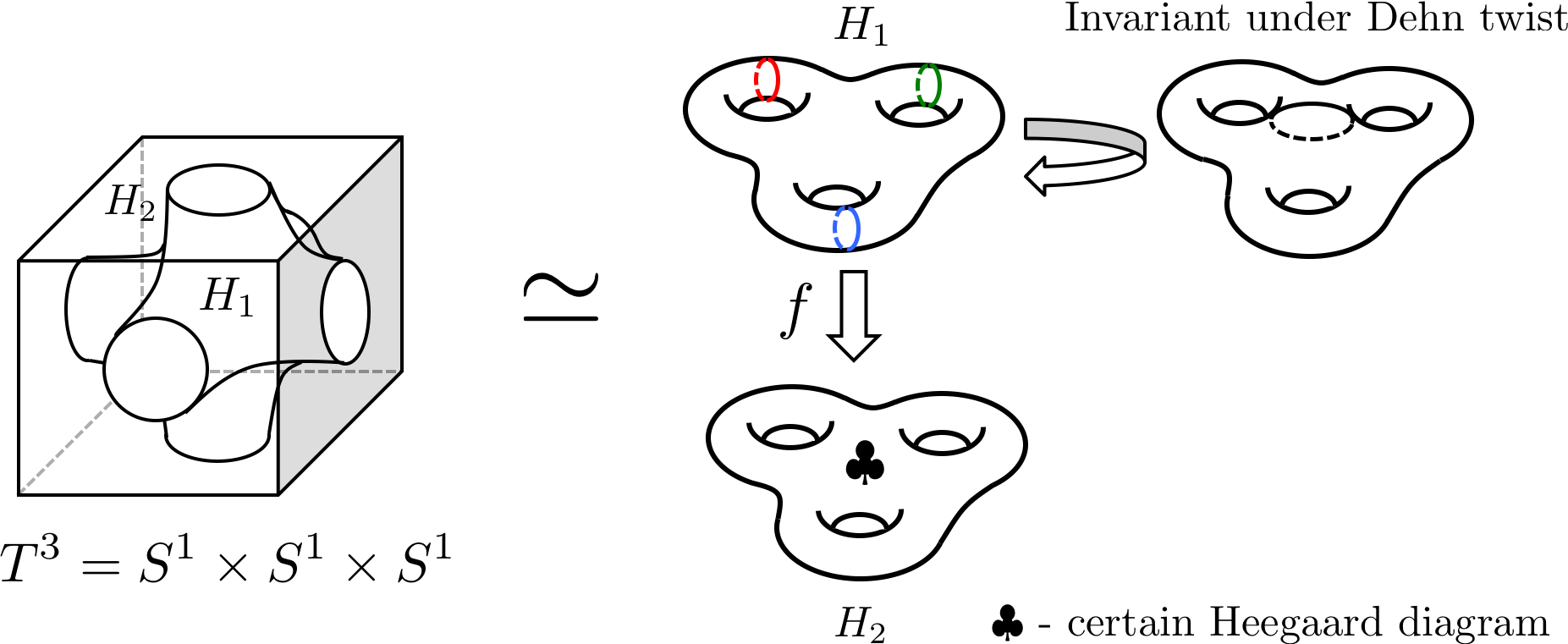}
\caption{A Heegaard decomposition of 3 torus. The explicit diagram of $\clubsuit$ is irrelevant to the proof.}
\label{ht3}
\end{figure}

For a 3 torus $T^3=S^1\times S^1 \times S^1$, it is convenient to represent it by a cube with opposite face identified. A Heegaard decomposition can be visualized in such a cube, see Fig.\ref{ht3}

The visualization provides a $g=3$ Heegaard decomposition. To construct the Heegaard diagram, we first pick a system of disks on the faces of cubes(in total 3, with opposite face identified). Its boundary produces a collection of curves on $\partial H_1$, as shown in Fig.\ref{ht3}. It also produces a collection of curves on the $\partial H_2$, as a `footprint' of the previous curves. We denote it by $\clubsuit$ in Fig.\ref{ht3}, the explicit diagram is complicated and not necessary for the proof.

We should mention that this specific decomposition of $T^3$ is different from the one in the main text, see Fig.\ref{dehn}. The reason we use a $g=3$ surface to do the decomposition here is to ensure the complement is also a handlebody, which is important to make the tools of Heegaard decomposition applicable in our proof. 

However, we will show this extra handle is irrelevant for the arguments here, that is to say, if we can prove (A) with the extra handle, we can also prove it without the extra handle in parallel. It is not true in general, but is true for the special surgery we are considering here. The reason is the following: a Dehn twist along a curve $c$ is a \textbf{local operation} on a tubular neighborhood A( will be define in \ref{Dehn twist}) on surface. This can be justified by providing a concrete definition of Dehn twist:

\begin{defn}
\label{Dehn twist}
Dehn twist:

Suppose $c$ is a simple closed curve on a closed orientable surface S. Let A be a tubular neighborhood of c, i.e., an annulus homeomorphic to $S^1\times I$, I is unit inverval. Now provide A a coordinates $(s,t)$, where $s$ is of the form $e^{i\theta}$ with $\theta\in[0,2\pi]$, and $t\in[0,1]$. Let $f$ be the map from S to itself, inside A, we require:

\begin{eqnarray}
f: (s,t) \mapsto (se^{i2\pi t},t)
\end{eqnarray}

outside A, $f$ is an identity map. Then we call $f$ a Dehn twist about the curve c.
\end{defn}

According to the definition, as long as we can move the extra handle far away from the tubular neighborhood A (which is possible in our case) the existence or absence of the extra handle won't change the effect of Dehn twist on the topology of $T^3$. Because the Dehn twist is an identity map outside A. By this argument, we relate the `surgery' in the main text to the one in this section, therefore, we just need to prove in the current setup, that relevant surgery won't change the topology of $T^3$.

On the other hand, it is logically possible to use $g=3$ handlebody to show that a crossing can be geometrically represented by a Dehn twist on Riemannian surface $\Sigma_3$ in the beginning, and prove the holding 3-d manifold $T^3$ is unchanged topologically. Also, the proof of (B) is for general $\Sigma_g$, therefore, is still applicable to show such crossing is in general changing spin structure, but a double crossing keep spin structure invariant. Nevertheless we choose the current presentation because it is obscure to use a $g=3$ surface in decomposition before knowing we need to prove some technical detail with a special tool: Heegaard decomposition.

Now to continue the proof (A) in the $g=3$ Heegaard decomposition, we need to identify the relevant `Dehn twist' (relevant to physical crossing of strings) on the handlebody picture and see its action on the diagrams.

\begin{figure}[htb]
\centering
\includegraphics[scale=0.2]{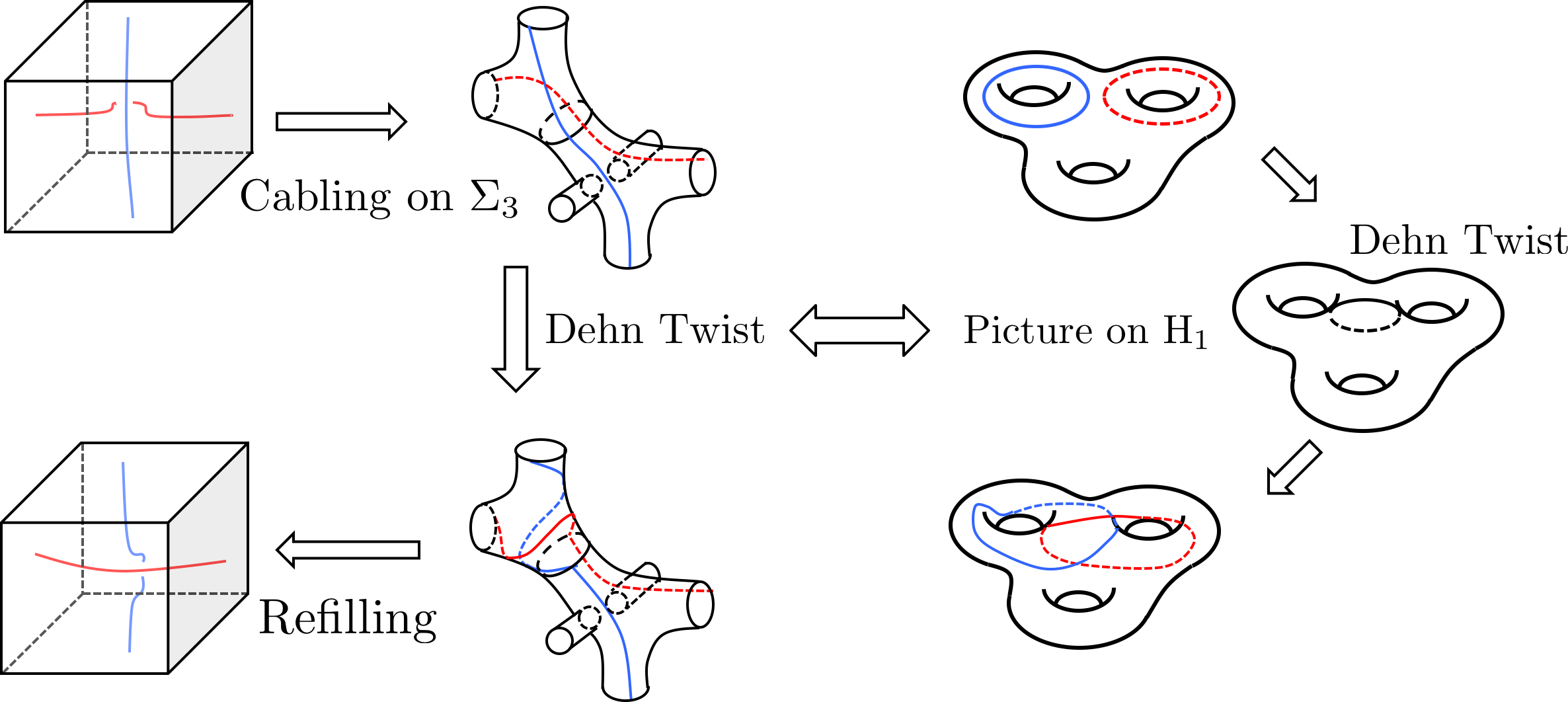}
\caption{Identifying the relevant Dehn twist. The relevant Dehn twist is around the curve between two top holes.}
\label{vis}
\end{figure}

With visualization shown in Fig.\ref{vis}, we can identity the relevant Dehn twist on the surface $\partial H_1$. It is around the curve between two top holes. Remarkably, this Dehn twist keeps the 3 circles we chosed on $\partial H_1$ (see colored circles in Fig.\ref{ht3}) unchanged. As a trivial corollary, these 3 circles are still qualified to be the boundary of a system of disks. Therefore after gluing back, we reproduce the same manifold $T^3$ topologically. 

\subsection{Proof of (B)}

In this subsection we present a proof of the statement (B): a double Dehn twist doesn't change spin structure. We will first describe the definition of spin structure in a precise language, state a classification theorem, and then give a computational proof.

\begin{defn}

(Haefliger(1956)\cite{haefliger1956lextension}):

A spin structure on an orientable Riemannian $n$ manifold $(M,g)$ is an equivariant lift of frame bundle $F_{SO}(M)\rightarrow M$, with respect to double covering $Spin(n)\rightarrow SO(n)$. I.e., a pair $(P,F_p)$ is a spin structure on the principal bundle $\pi$: $F_{SO} \rightarrow M$, if:

(a) $\pi_P: P\rightarrow M$ is a principal spin bundle;

(b) $F_p$ is a equivariant 2-fold covering map.
\end{defn}

Physically speaking, a spin structure is a consistent choice of periodic/anti-periodic boundary conditions for fermions, similar but different to a `$\mathbb{Z}_2$ gauge field'. (Different in the way that there is a special trivial configuration for gauge field, $i.e.$, all holonomy are trivial. While for spin structure, generically, there is no pre-selected `trivial' configuration.)

Remarkably, there is a classification theorem of spin structure(e.g., see Ref.\cite{atiyah1971riemann}\cite{johnson1980spin}):

\begin{thm}
the set of spin structures on a closed surface $\Sigma_g$ of genus g can be identified with the set of quadratic forms on $H_1(\Sigma_g;\mathbb{Z}_2)$
\end{thm}

Now we will calculate the action of a double Dehn twist on the mod 2 homology $H_1(\Sigma_g;\mathbb{Z}_2)$.

A Dehn twist $t_a$ along a simple closed curve $a\subset \Sigma_g $ acts on mod 2 homology $x\in H_1(\Sigma_g,\mathbb{Z}_2)$ in the following way:

\begin{eqnarray}
t_a(x)=x+(a\cdot x) \alpha
\end{eqnarray}
where $\alpha$ are the homology class for $a$, $a\cdot x$ counts the mod 2 intersection points between a representative of class $x$ and a. Therefore, a double of Dehn twist:

\begin{eqnarray}
t_a^2(x)&=&(x+(a\cdot x) \alpha)+ (a\cdot (x+(a\cdot x) \alpha)) \alpha\nonumber \\
&=&x + 2 (a\cdot x) \alpha = x \qquad \text{Mod 2}
\end{eqnarray}
acts trivially on mod 2 homology. An immediate corollary says it preserves the spin structure.

\subsection{Final subtlety}
\label{final}

In the main text, for simplicity, we model $2\pi$ flux tube by featureless lines. This might be unrealistic because the real physical flux tube always has a finite size $\sim \lambda_p$, the penetration depth. So in general, we should consider finer structure associated with those lines: the framing. Pictorially, it is sufficient to use a ribbon with one solid side and one dash side to represent the framing structure. We always imagine an arrow from the solid side to the dash side representing the framing.

\begin{figure}[h]
\center
\includegraphics[scale=0.3]{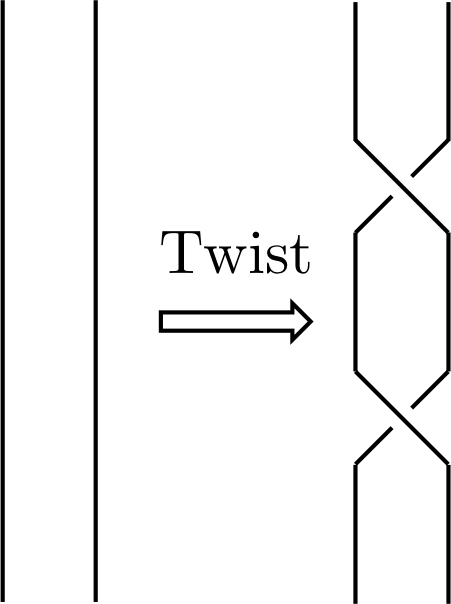} \qquad \qquad \includegraphics[scale=0.3]{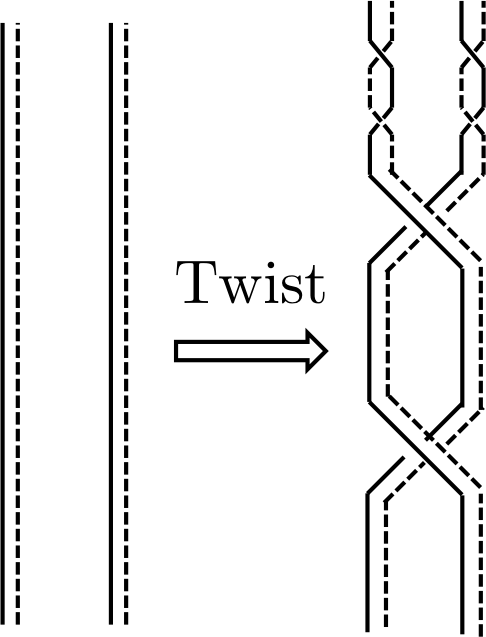}
\caption{Twist (same effect as Dehn twist acting on the surface these lines cabled on) for two lines and two ribbons. Leftside is a twist for line, rightside is a twist for ribbon: it will produce a self twist for each ribbon as a by-product.}
\label{twist}
\end{figure}

An important subtlety in ribbon case is the effect of a twist on two ribbons together, see Fig.\ref{twist}. In the `line' case, we can conveniently represent a crossing by a twist locally, e.g., after cabling these two lines on a surface, a twist can be realized as a Dehn twist along a curve intersecting these two lines. However, for ribbon case, a twist of two ribbons will induce a self twist for each ribbon. (For a similar discussion, see Ref.\cite{freedman2011projective}.)

\begin{figure}[h]
\center
\includegraphics[scale=0.3]{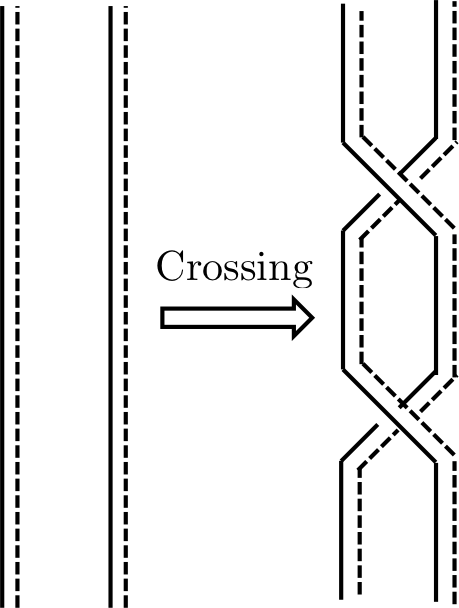}
\caption{A pure crossing for ribbons could be viewed as a total twist for two ribbons together with an anti-twist for each ribbon individually $C_{ab}=T^{-1}_{a} T^{-1}_b T_{ab}$.}
\label{Crossing ribbon}
\end{figure}

Therefore, to represent a pure crossing for ribbons, we should undo the twist for each ribbon after a total twist for two ribbon together see Fig.\ref{Crossing ribbon}. If we denote left ribbon by $a$, and right ribbon by $b$, then, a crossing between a and b $C_{ab}$ can be expressed as an algebraic formula: $C_{ab}=T^{-1}_{a} T^{-1}_b T_{ab}$. T stands for a twist.

\begin{figure}[h]
\center
\includegraphics[scale=0.3]{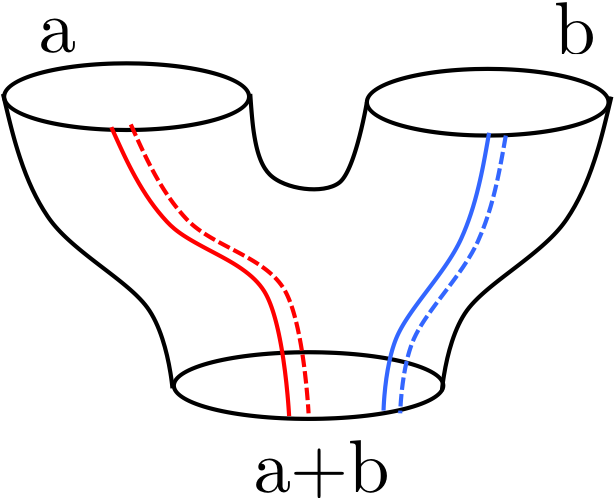}
\caption{a crossing for a pair of ribbons $(a,b)$ could be modeled as a combination of Dehn twist along circle $a+b$ and anti Dehn twist along circle $a$ and $b$ individually}
\label{crossing dehn}
\end{figure}

Taking this subtlety in count, a crossing for a pair of $2\pi$ flux in the main text could be essentially modeled as a combination of Dehn twist along circle $a+b$ and anti Dehn twist along circle $a$ and $b$ individually, see Fig.\ref{crossing dehn}. It is straightforward to check that all these three Dehn twists still keep the Heegaard diagram invariant, therefore, the topology of hosting manifold is still $T^3$ after crossing.

Now we discuss the modification to the statement (B) with ribbon structure. Again, we can write down the action of Dehn twists on mod 2 homology, denote the combination by $c_{ab}=t^{-1}_a t^{-1}_b t_{a+b}$, and notice $a$ and $b$ do not intersect:

\begin{eqnarray}
c_{ab}(x) &=&t^{-1}_a t^{-1}_b t_{a+b} (x)\\
&=& t^{-1}_a t^{-1}_b (x+((a+b)\cdot x)(\alpha+\beta))\\
&=& t^{-1}_a (x+((a+b)\cdot x)(\alpha+\beta)-(b\cdot x)\beta ) \\
&=& x+((a+b)\cdot x)(\alpha+\beta)\\&-&(b\cdot x)\beta-(a\cdot x)\alpha\\
&=&x+(a\cdot x)\beta+(b\cdot x)\alpha
\end{eqnarray}

So a double of crossing:

\begin{eqnarray}
c_{ab}^2(x) &=& c_{ab} (x+(a\cdot x)\beta+(b\cdot x)\alpha) \\
&=& x \ \text{(mod 2)}
\end{eqnarray}
acts trivially on mod 2 homology as well. This finishes the proof in ribbon case.

\section{Some basics of Clifford algebra and its representations}

\label{Clifford}

This section is devoted to prepairing necessary tools about Clifford algebra for section(\ref{higherC}) and section(\ref{fNLSM}). Although all results in this section are well-known in the literature, we would like to include this background knowledge for completeness.

To begin with, we fix the notation: 

\begin{defn}

Clifford algebra $Cl(p,q)$ is generated by a vector space V with basis $\lbrace e_1, e_2, \ldots e_{p+q} \rbrace$, subject to conditions:

\begin{eqnarray}
\lbrace e_i, e_j \rbrace &=&0  \qquad i\neq j \\
e_i^2 &=& 1 \qquad i=1, \ldots, p \\
e_i^2 &=& -1\ \quad i=p+1, \ldots, p+q
\end{eqnarray}

Clifford algebra $Cl(p,q)$, as a vector space over field $\mathbb{K}$, (e.g., $\mathbb{K}=\mathbb{R}$ or $\mathbb{C}$), has dimension $2^n$ ($n=p+q$), is freely generated by $2^{n}$  basis vectors:

\begin{eqnarray}
e_{i_1} e_{i_2} \ldots e_{i_k} \qquad i_1 < \ldots < i_k, \qquad 1\leq k \leq n
\end{eqnarray}

\end{defn}

Among all the elements, there is one element of particular importance, often called chiral element:

\begin{defn}
Chiral element:
\begin{eqnarray}
e_{c}=e_1 e_2 e_3 \ldots e_n
\end{eqnarray}
When n is odd, $e_c$ commutes with every generator, so $Cl(2k+1)$ has nontrivial center spanned by 1 and $e_c$. For $Cl(2k)$, the center is trivial (only span by 1).
\end{defn}

For physical application, we restrict ourselves to real and complex numbers, $i.e.,$ $\mathbb{K}=\mathbb{R}$ or $\mathbb{C}$. Then Clifford algebra over $\mathbb{C}$ can be understood as complex extension of real one, and will be denoted as $Cl^c$:

\begin{eqnarray}
Cl^c(n)=Cl(n) \otimes_{\mathbb{R}} \mathbb{C}
\end{eqnarray}
Since $\mathbb{C}$ includes the square root of $-1$, there will be no difference between positive and negative generators:

\begin{eqnarray}
Cl^c(p+q)=Cl(p,q) \otimes_{\mathbb{R}} \mathbb{C}
\end{eqnarray}

We are going to discuss the representations of Clifford algebra over $\mathbb{R}$ and $\mathbb{C}$. As a starting point of our discussion, it is useful to notice the theorem by Wedderburn: 

\begin{thm}
Wedderburn theorem:

Any simple algebra over $\mathbb{R}$ or $\mathbb{C}$ is isomorphic to certain matrix algebra over the corresponding division algebra. 
\end{thm}
According to Frobenius theorem:
\begin{thm}
Frobenius theorem:

the only division algebras over $\mathbb{R}$ are real $\mathbb{R}$, complex $\mathbb{C}$ and quaternion $\mathbb{H}$; the only division algebra over $\mathbb{C}$ is $\mathbb{C}$.

\end{thm}

These two theorems allow us to obtain a general classification about the Clifford algebra of our interest. We will discuss the details in the following subsections.

\begin{table*}[t]
\caption{Real Clifford algebra $Cl(p(+),q(-))$}
\begin{tabular}{c |c c c c c c c c}
$(p,q)$ &0 & 1 & 2 & 3 & 4& 5 & 6 & 7 \\
\hline
0 & $\mathbb{R}$ & $\mathbb{C}$  & $\mathbb{H}$ & $\mathbb{H}\oplus \mathbb{H}$ & $\mathbb{H}(2)$ & $\mathbb{C}(4)$ & $\mathbb{R}(8)$ & $\mathbb{R}(8)\oplus \mathbb{R}(8)$ \\
1 & $\mathbb{R}\oplus \mathbb{R}$   & $\mathbb{R}(2)$ & $\mathbb{C}(2)$  & $\mathbb{H}(2)$ & $\mathbb{H}(2)\oplus \mathbb{H}(2)$ & $\mathbb{H}(4)$ & $\mathbb{C}(8)$ & $\mathbb{R}(16)$ \\
2 & $\mathbb{R}(2)$ &$\mathbb{R}(2)\oplus \mathbb{R}(2)$   & $\mathbb{R}(4)$ & $\mathbb{C}(4)$  & $\mathbb{H}(4)$ & $\mathbb{H}(4)\oplus \mathbb{H}(4)$ & $\mathbb{H}(8)$ & $\mathbb{C}(16)$  \\
3 & $\mathbb{C}(2)$   & $\mathbb{R}(4)$ &$\mathbb{R}(4)\oplus \mathbb{R}(4)$   & $\mathbb{R}(8)$ & $\mathbb{C}(8)$  & $\mathbb{H}(8)$ & $\mathbb{H}(8)\oplus \mathbb{H}(8)$ & $\mathbb{H}(16)$ \\
4 & $\mathbb{H}(2)$ & $\mathbb{C}(4)$   & $\mathbb{R}(8)$ &$\mathbb{R}(8)\oplus \mathbb{R}(8)$   & $\mathbb{R}(16)$ & $\mathbb{C}(16)$  & $\mathbb{H}(16)$ & $\mathbb{H}(16)\oplus \mathbb{H}(16)$  \\
5 & $\mathbb{H}(2)\oplus \mathbb{H}(2)$ & $\mathbb{H}(4)$ & $\mathbb{C}(8)$   & $\mathbb{R}(16)$ &$\mathbb{R}(16)\oplus \mathbb{R}(16)$   & $\mathbb{R}(32)$ & $\mathbb{C}(32)$  & $\mathbb{H}(32)$  \\
6  & $\mathbb{H}(4)$ & $\mathbb{H}(4)\oplus \mathbb{H}(4)$ & $\mathbb{H}(8)$ & $\mathbb{C}(16)$   & $\mathbb{R}(32)$ &$\mathbb{R}(32)\oplus \mathbb{R}(32)$   & $\mathbb{R}(64)$ & $\mathbb{C}(64)$   \\
7  & $\mathbb{C}(8)$    & $\mathbb{H}(8)$ & $\mathbb{H}(8)\oplus \mathbb{H}(8)$ & $\mathbb{H}(16)$ & $\mathbb{C}(32)$   & $\mathbb{R}(64)$ &$\mathbb{R}(64)\oplus \mathbb{R}(64)$   & $\mathbb{R}(128)$ 
\end{tabular}
\label{R}
\end{table*}

\subsection{Structure and representation of complex Clifford algebra}

Let's first do the easier case: complex Clifford algebra $Cl^c(n)$. As we have seen in the above discussion, the center of $Cl^c(n)$ depends on the parity of n: $Cl^c(2k)$ has trivial center, so it is isomorphic to complex matrix algebra $\mathbb{C}(2^k)$(dimension is determined by counting $2^{2k}=(2^k)^2$):

\begin{eqnarray}
Cl^c(2k) \cong \mathbb{C}(2^k)
\end{eqnarray}
More explicitly, there are $2k$ gamma matrices with dimension $2^k$, $\gamma_i$ as a representation of generators, $e_i,\ (i=1,\ldots, 2k)$. 

For the $Cl^c(2k+1)$ case, the center is nontrivial, generated by 1 and $e_c$, so the algebra $Cl^c(2k+1)$ has to be a direct sum of two matrix algebra $\mathbb{C}(2^k)$:

\begin{eqnarray}
Cl^c(2k+1) \cong \mathbb{C}(2^k) \oplus \mathbb{C}(2^k)
\end{eqnarray}
this also agrees with the dimension counting: $2^{2k+1}=(2^k)^2\times 2$. In terms of representation, the two parts can be understood as two different representations for the chiral element. For example, $Cl(3)$ has two representations, Pauli matrices $\sigma_x,\sigma_y,\sigma_z$ and its complex conjugation, corresponding to $e_c=i$ and $-i$.

More generally, for $Cl(2k+1)$, the square of chiral element $e_c^2=(-1)^{k(2k+1)}$. The two representations correspond to $e_c= \pm 1$ for $k$ even, and $e_c=\pm i$ for $k$ odd.

\subsection{Representation of real Clifford algebra}

The more subtle case is the real Clifford algebra, because we have to pay attention to the difference between positive and negative generators. We will keep the original notation and denote Clifford algebra with signature $\lbrace p(+), q(-)  \rbrace$ as $Cl(p,q)$.

According to the general principle stated at beginning, the real Clifford algebra is isomorphic to matrix algebra over $\mathbb{K}=\mathbb{R},\ \mathbb{C}$ or $\mathbb{H}$. The matrix algebra has obvious properties:

\begin{eqnarray}
\mathbb{K}(n) &\cong & \mathbb{R} (n) \otimes_{\mathbb{R}} \mathbb{K} \label{a1}\\
\mathbb{R} (nm) &\cong & \mathbb{R}(n) \otimes_{\mathbb{R}} \mathbb{R} (m)  \label{a2}
\end{eqnarray}

Among these three types, $\mathbb{R}(n)$ and $\mathbb{H}(n)$ are central algebra ($i.e.$, have trivial center). While $\mathbb{C}(n)$ is not a central algebra, $i=\sqrt{-1}$ is a nontrivial center. Remember that  the center of $Cl(p,q)$ is trivial when $p+q$ is even, and nontrivial when $p+q$ is odd, we conclude that

\begin{eqnarray}
Cl(p,2n-p) &\cong& \mathbb{R}(m) \ or \  \mathbb{H}(m) \\
Cl(p,2n+1-p) &\cong& \mathbb{R}(m) \oplus \mathbb{R}(m) \nonumber\\
 &or&   \mathbb{H}(m) \oplus \mathbb{H}(m)\nonumber \\
 &or&  \mathbb{C}(m)
\end{eqnarray}

To get a finer resolution, we square the chiral element in $p+q=2k+1$ case. Using the anti-commutation relations of the generators one obtains
\begin{eqnarray}
e_c^2=(-1)^{\frac{n(n-1)}{2}+q}
\end{eqnarray}
When $e_c^2=-1$, the algebra contains a complex structure induced by $e_c$.

The above are some general structures which can be verified from the result that we are going to derive now. The following construction is following the strategy of Ref.\cite{atiyah1964clifford}. Let us first state the `rules' and then give a quick constructive proof.

There is a way to construct larger Clifford algebra by smaller one, following the rules:

\begin{eqnarray}
Cl(p,q) \otimes Cl(2,0) &\simeq & Cl(q+2,p) \\
Cl(p,q) \otimes Cl(0,2) &\simeq & Cl(q,p+2) \\
Cl(p,q) \otimes Cl(1,1) &\simeq & Cl(p+1,q+1)
\end{eqnarray}

\begin{pf}
Here we provide a constructive proof, following which, we can construct right hand side(RHS) from the data of left hand side(LHS).

Generators for the LHS:
\begin{eqnarray}
\lbrace e_1 \ldots e_p , e_{p+1}, \ldots , e_{p+q}  \rbrace &\otimes& \lbrace e^\prime_1,e^\prime_2 \rbrace
\end{eqnarray}
First, produce two generators of RHS simply by tensor with Identity:
\begin{eqnarray}
 \lbrace  \mathbb{I}\otimes e^\prime_1,   \mathbb{I}\otimes e^\prime_2 \rbrace
\end{eqnarray}
Second, the rest of the generators are constructed by tensor of the generators from $Cl(p,q)$ with $e'\equiv e^\prime_1 e^\prime_2$:
\begin{eqnarray}
\lbrace e_1\otimes e^\prime, \ldots e_p\otimes e^\prime,e_{p+1}\otimes e^\prime, \ldots , e_{p+q}\otimes e^\prime \rbrace
\end{eqnarray}
It is straightforward to verify the above constructions do satisfy the rules for the generators of RHS.

\end{pf}

The above rules provide a way to construct general $Cl(p,q)$ from building blocks. The building blocks are in low dimensions, which is simple enough to be figured out case by case:

\begin{eqnarray}
Cl(1,0) &\cong& \mathbb{R} \oplus \mathbb{R} \\
Cl(0,1) &\cong& \mathbb{C}\\
Cl(2,0) &\cong& \mathbb{R}(2)\\
Cl(0,2) &\cong& \mathbb{H}
\end{eqnarray}

Another ingredient from matrix algebra are the `tensor rules' over $\mathbb{R}$:
\begin{eqnarray}
\mathbb{C} \otimes_{\mathbb{R}} \mathbb{C} &\cong& \mathbb{C} \oplus \mathbb{C}\\
\mathbb{C} \otimes_{\mathbb{R}} \mathbb{H} &\cong& \mathbb{C}(2)\\
\mathbb{H} \otimes_{\mathbb{R}} \mathbb{H} &\cong& \textrm{Hom}_{\mathbb{R}}(\mathbb{H}, \mathbb{H}) \cong \mathbb{R}(4)
\end{eqnarray}
Following these receipts, we can build up the whole table within a Bott period. See Table(\ref{R}). 

From the table, we can read out the minimal dimension of their real representations. Physically speaking, it is the minimal dimension of spinor representations. We will use $dim_R(\mathbb{K}(n))$ to denote the dimension of the corresponding spinors/representation spaces (not the dimension of the algebra itself). $dim_R(\mathbb{K}(n))$ follows the rules:
\begin{eqnarray}
dim_{\mathbb{R}}(\mathbb{K}(n))&=&n \times  dim_{\mathbb{R}}(\mathbb{K})\\
dim_{\mathbb{R}}(\mathbb{R})&=&1\\
dim_{\mathbb{R}}(\mathbb{C})&=&2\\
dim_{\mathbb{R}}(\mathbb{H})&=&4
\end{eqnarray}

\section{A free fermion computation of higher Chern numbers}
\label{higherC}

This section devotes to a computation of higher Chern numbers used in the main text.

For non-interacting gapped fermions in $(4n+1)$ space-time dimensions, the Chern number $c_{2n}$ determines the number of chiral edge states. More explicitly, a bulk theory with Chern number $c_{2n}$ has an edge of $|c_{2n}|$ copy(s) of $4n$-d chiral fermions, with the chirality determined by the sign of $c_{2n}$ ({\it c.f.} Ref\cite{qi2008topological}). The unit of this quantization may change when additional symmetries are enforced. Our strategy to compute the quantization of Chern number is by realizing the edge Hamiltonian preserving symmetries, and check how many copies of chiral fermions we need in the minimal realization. In order to implement the symmetries (including anti-unitary symmetries), we follow Kitaev's strategy\citep{kitaev2009periodic} and write down the surface state Hamiltonian:

\begin{eqnarray}
H= \frac{i}{2} \int d^4 k   \psi^T \Gamma^j \partial_j \psi
\end{eqnarray}
Here $\Gamma^j$ are real gamma matrices, and both time reversal T, and ${\rm U(1)}$ symmetry Q can be realized as unitary matrices.
\begin{eqnarray}
T^2=Q^2= -1 &\quad &  TQ=-QT \\
\left[ H,Q \right]=0 &\quad & \lbrace H,T \rbrace =0
\end{eqnarray}

To fit into Clifford algebra, we consider combination of T and Q, $e=TQ$, which anti-commutes with Hamiltonian and T, and squared to be $-1$. Therefore, $\lbrace \Gamma^j, T, e=TQ \rbrace$ forms a set of generators of $Cl(3,2)$. According to the Table(\ref{R}):

\begin{eqnarray}
Cl(3,2) &\cong & \mathbb{R}(4) \oplus \mathbb{R}(4) \\
dim_{\mathbb{R}}(3,2) &\equiv& dim_{\mathbb{R}}(Cl(3,2))= 4
\end{eqnarray}
The dimension of spinors on chiral edge of $4+1$d Chern insulator would be $4$. Here is an example of explicit realization:
\begin{eqnarray}
\gamma_1 &=& \sigma_1 \otimes \mathbb{I} \quad
\gamma_2 = \sigma_2 \otimes \sigma_2 \quad
\gamma_3 = \sigma_3 \otimes \mathbb{I} \\
T &=& i \sigma_2 \otimes \sigma_1 \quad
TQ = i \sigma_2 \otimes \sigma_3
\end{eqnarray}

In a similar way, the minimal $(3+1)$-d chiral fermion (which is required to have symmetry $Q$ but not necessarily $T$, so we apply complex Clifford algebra) has the dimension $dim_{\mathbb{R}}(Cl^c(3))=4$. Therefore, imposing time-reversal symmetry (with $T^2=-1$) does not require a duplication of surface fermion chirality, which implies that the unit of Chern number for $T^2=-1$ symmetry is $c_2=1$.

As a comparison, if we require a time-reversal symmetry with $T^2=1$, then T and TQ are positive generators in the Clifford algebra, so that the algebraic structure for this $Cl(5,0)$ would be different:
\begin{eqnarray}
Cl(5,0) &\cong& \mathbb{H}(2) \oplus \mathbb{H}(2) \\
dim_{\mathbb{R}}(5,0)&=&8
\end{eqnarray}
which means in order to get the correct algebra for the edge states, we have to double the number of chiral fermions. This implies that the quantization unit for Chern number $c'_2$ here is doubled: $c'_2=2n,~n\in \mathbb{Z}$.

Now let us consider the surface state of $(8+1)$-d Chern insulator with $T^2=-1$, which corresponds to 
\begin{eqnarray}
Cl(7,2)&\cong &  \mathbb{H}(8) \oplus \mathbb{H}(8)  \\
dim_{\mathbb{R}}(7,2) &=& 32
\end{eqnarray}
On comparison, the `standard' chiral fermion without enforcement of time reversal symmetry, has minimal spinor dimension 16: 
\begin{eqnarray}
Cl^c(7) &\cong& \mathbb{C}(8)\oplus \mathbb{C}(8) \\
dim_\mathbb{R}(Cl^c(7))&=&16
\end{eqnarray}
This implies that we have to double the number of chiral fermions in order to incorporate time-reversal symmetry, which means the quantization for Chern number $c_4$ is doubled. (As a side remark, if we consider time-reversal symmetry with $T^2=1$ instead of $-1$, we obtain $Cl(9,0)\cong \mathbb{R}(16)$ so that the minimal unit for $c'_4$ is: $c'_4=1$. )

The Bott periodicity asserts that the two examples we check determine the situation in all higher $4n$ dimensions: with $T^2=-1$ symmetry, the minimal unit is $c_{2n}=1$ when $n$ is odd, and $c_{2n}=2$ when $n$ is even.

\section{Comparison to fermion $\sigma$ model approach}
\label{fNLSM}

In this section, we study the topological classification of DIII TSC in $4n$-dimensions with the fermion $\sigma$ model approach\citep{Kitaev}\cite{you2014symmetry}.

We consider a generic fermion model on spacetime dimension $4n-1$ (the boundary of $4n$-d TSC) with N mass terms:
\begin{eqnarray}
H= i \int d^{4n-2}x\  \psi^T \left(  - \Gamma^j \partial_j + \gamma^a m_a \right) \psi
\end{eqnarray}
where we factor the pure imaginary number $i$ out, and use Majorana representation\cite{kitaev2009periodic}. Gamma matrices $\Gamma^i, \ i=1,\ldots, 4n-2 $ and $\gamma_a, \ a=1, \ldots, N$ are positive and negative generators of $Cl(4n-2,N)$ respectively.

Following the notation from section(\ref{Clifford}), the dimension for  spinors in massive model is $dim_{\mathbb{R}}(4n-2,N)$, while the dimension for massless spinors with T symmetry is $dim_{\mathbb{R}}(4n-2,1)$. Therefore the number of copies needed to carry $N$ T-breaking mass terms is:
\begin{eqnarray}
\nu =\frac{dim_{\mathbb{R}}(4n-2,N)}{dim_{\mathbb{R}}(4n-2,1)}
\end{eqnarray}
By integrating out the fermion modes, we get an $O(N)$ $\sigma$ model with the target space a sphere $S^{N-1}$. The critical dimension below which such a $\sigma$ model doesn't have any topological term is space-time dimension $N-2$, where a Weiss-Zumino-Witten (WZW) term\cite{wess1971,witten1983,witten1984} is defined. That is to say, if $N-2>4n-1$, the $\sigma$ model can not have any topological term, so that the mass order parameter could be safely disordered without closing the gap.

According to this argument, we obtain the minimal number of copies needed to trivialize the surface theory, by setting $N=4n+2$:
\begin{eqnarray}
\nu =\frac{dim_{\mathbb{R}}(4n-2,4n+2)}{dim_{\mathbb{R}}(4n-2,1)}
\end{eqnarray}

Following the procedure described in section(\ref{Clifford}), we can compute $\nu$ explicitly:

\begin{itemize}
\item For space-time dimension $d+1=4n$ with odd $n$: $n=2k+1$, $\nu =2^{4k+4}=2^{\frac{d+5}2}$ 

\item For space-time dimension $d+1=4n$ with even $n$: $n=2k$, $\nu =2^{4k+1}=2^{\frac{d+3}2}$
\end{itemize}
This result agrees with the axion field theory result in the main text. 

We would like to finish this section by a comment about the relation between the $\sigma$ model approach and the axion field theory approach. In the $\sigma$ model approach, it is relatively easy to understand why $\nu$ copies is {\it sufficient} to trivialize the surface, because there is nothing to forbid a disorder phase when we have enough ($N>4n+1$) mass terms. On the other hand, it is generally hard to demonstrate whether TSC with less than $\nu$ copies of surface states is indeed topologically {\it nontrivial}. When the number of mass terms $N\leq 4n+1$, there might be topological terms in the $\sigma$ model, ($e.g.,$ WZW term when $N=4n+1$) but to prove that a trivial disordered phase does not exist, one needs to consider the possibility of more complicated mass terms corresponding to target manifolds that are not spheres. In contrast, the axion field theory approach explicitly shows that any TSC with the topological invariant nonzero and less than $\nu$ is nontrivial, due to the anomaly,  but it does not directly predict that TSC with topological invariant equal to $\nu$ is trivial. In this sense, combining the two arguments, we are able to completely pin down the classification for general $4n$ dimensional DIII class.

\end{document}